\documentclass[preprint2]{aastex}

\slugcomment{ApJ accepted}

\shorttitle{SEDs of AGNs in the XMDS Survey}
\shortauthors{Polletta et al.}

\def\nodata{{...}}

\def\rp{$r^\prime$}
\def\gp{$g^\prime$}
\def\rp{$r^\prime$}
\def\ip{$i^\prime$}

\def\lsun{L$_{\odot}$}

\def\chandra {{\it Chandra}}
\def\xmm {XMM-{\it Newton}}
\def\spitzer {{\it Spitzer}}
\def\nh {${\rm N_\mathrm{H}}$}
\def\av {${\rm A_\mathrm{V}}$}

\def\micron{$\mu$m}
\def\kms{\ifmmode {\rm\,km\,s^{-1}}\else
    ${\rm\,km\,s^{-1}}$\fi}
\def\kmsMpc{\ifmmode {\rm\,km\,s^{-1}\,Mpc^{-1}}\else
    ${\rm\,km\,s^{-1}\,Mpc^{-1}}$\fi}
\def\ergAcm2{\ifmmode {\rm\,erg\,cm^{-2}\,{\rm \AA}^{-1}}\else
    ${\rm\,erg\,cm^{-2}\,\AA^{-1}}$\fi}
\def\cm2{\ifmmode {\rm\,cm^{-2}}\else
    ${\rm\,cm^{-2}}$\fi}
\def\ergcm2s{\ifmmode {\rm\,erg\,cm^{-2}\,s^{-1}}\else
    ${\rm\,erg\,cm^{-2}\,s^{-1}}$\fi}
\def\cgsdeg2{\ifmmode {\rm\,erg\,cm^{-2}\,s^{-1}\,deg^{-2}}\else
    ${\rm\,erg\,cm^{-2}\,s^{-1}\,deg^{-2}}$\fi}
\def\sqdeg{\ifmmode {\rm\,deg^{2}}\else
    ${\rm\,deg^{2}}$\fi}
\def\ergsHz{\ifmmode {\rm\,erg\,s^{-1}\,Hz^{-1}}\else
    ${\rm\,erg\,s^{-1}\,Hz^{-1}}$\fi}
\def\ergs{\ifmmode {\rm\,erg\,s^{-1}}\else
    ${\rm\,erg\,s^{-1}}$\fi}
\def\ergsA{\ifmmode {\rm\,erg\,s^{-1}\,\AA^{-1}}\else
    ${\rm\,erg\,s^{-1}\,\AA^{-1}}$\fi}
\def\WHz{\ifmmode {\rm\,W\,Hz^{-1}}\else
    ${\rm\,W\,Hz^{-1}}$\fi}
\def\WHzsr{\ifmmode {\rm\,W\,Hz^{-1}\,sr^{-1}}\else
    ${\rm\,W\,Hz^{-1}\,sr^{-1}}$\fi}
\def\ergscm2Hz{\ifmmode {\rm\,erg\,cm^{-2}\,s^{-1}\,Hz^{-1}}\else
    ${\rm\,erg\,cm^{-2}\,s^{-1}\,Hz^{-1}}$\fi}

\def\oiii{[\ion{O}{3}] $\lambda$5007}

\begin{document}


\title{Spectral Energy Distributions of Hard X-ray selected AGNs in the XMDS Survey}


\author{M. Polletta\altaffilmark{1}, M. Tajer\altaffilmark{2,3}, L.
Maraschi\altaffilmark{2}, G. Trinchieri\altaffilmark{2}, C.J.
Lonsdale\altaffilmark{1}, L. Chiappetti\altaffilmark{4} S.
Andreon\altaffilmark{2}, M. Pierre\altaffilmark{5}, O. Le
F\`evre\altaffilmark{6}, G. Zamorani\altaffilmark{7}, D.
Maccagni\altaffilmark{4}, O. Garcet\altaffilmark{8}, J.
Surdej\altaffilmark{8}, A. Franceschini\altaffilmark{9}, D.
Alloin\altaffilmark{5}, D. L. Shupe\altaffilmark{10}, J. A.
Surace\altaffilmark{10}, F. Fang\altaffilmark{10}, M.
Rowan-Robinson\altaffilmark{11}, H. E. Smith\altaffilmark{1}, L.
Tresse\altaffilmark{6}}


\altaffiltext{1}{Center for Astrophysics \& Space Sciences, University of California, San Diego, La Jolla, CA  92093--0424, USA}
\altaffiltext{2}{INAF -- Osservatorio di Brera, via Brera 28, 20121 Milano, Italy}
\altaffiltext{3}{Universit\`a degli studi di Milano - Bicocca, Dipartimento di fisica, Piazza della Scienza 3, 20126 Milano, Italy}
\altaffiltext{4}{INAF -- IASF Milano, via Bassini 15, I-20133 Milano, Italy}
\altaffiltext{5}{CEA Saclay, DSM/DAPNIA, Service d'Astrophysique, 91191 Gif-sur-Yvette, France }
\altaffiltext{6}{Laboratoire d'Astrophysique de Marseille, UMR 6110 CNRS-Universit\'e de Provence, Traverse du Siphon, BP 8, 13012 Marseille, France}
\altaffiltext{7}{INAF -- Osservatorio di Bologna, via Ranzani 1, 40127 Bologna, Italy}
\altaffiltext{8}{Institut d'Astrophysique et de G\'eophysique, Universit\'e de Li\`ege, All\'ee du 6 Ao\^ut 17, 4000 Li\`ege 1, Belgium}
\altaffiltext{9}{Dipartimento di Astronomia, Universit\`a di Padova, Vicolo dell'Osservatorio 2, I-35122, Padova, Italy}
\altaffiltext{10}{\spitzer\ Science Center, California Institute of Technology, 100-22, Pasadena, CA 91125, USA}
\altaffiltext{11}{Astrophysics Group, Blackett Laboratory, Imperial College, Prince Consort Road, London, SW7 2BW, UK}


\begin{abstract}
We present the spectral energy distributions (SEDs) of a hard X-ray selected
sample. The sample contains 136 sources with F$_{2-10 keV}$$>$10$^{-14}$
\ergcm2s\ and 132 are AGNs. The sources are detected in a 1 deg$^2$ area of the
\xmm-Medium Deep Survey where optical data from the VVDS, CFHTLS surveys,
and infrared data from the SWIRE survey are available. Based on a SED 
fitting technique we derive photometric redshifts with
$\sigma$(1+$z$)=0.11 and 6\% of outliers and identify AGN signatures in
83\% of the objects. This fraction is higher than derived when a
spectroscopic classification is available. The remaining 17$^{+9}_{-6}$\% of
AGNs shows star-forming galaxy SEDs (SF class). The sources with AGN
signatures are divided in two classes, AGN1 (33$^{+6}_{-1}$\%) and AGN2
(50$^{+6}_{-11}$\%). The AGN1 and AGN2 classes include sources whose SEDs
are fitted by type 1 and type 2 AGN templates, respectively. On average,
AGN1s show soft X-ray spectra, consistent with being unabsorbed, while AGN2s
and SFs show hard X-ray spectra, consistent with being absorbed. The
analysis of the average SEDs as a function of X-ray luminosity shows a
reddening of the IR SEDs, consistent with a decreasing contribution from the
host galaxy at higher luminosities. The AGNs in the SF classes are likely
obscured in the mid-infrared, as suggested by their low $L_{3-20\mu
m}/L^{corr}_{0.5-10 keV}$ ratios. We confirm the previously found
correlation for AGNs between the radio luminosity and the X-ray and the
mid-infrared luminosities. The X-ray-radio correlation can be used to
identify heavily absorbed AGNs. However, the estimated radio fluxes for the
missing AGN population responsible for the bulk of the background at
$E>$10~keV are too faint to be detected even in the deepest current radio
surveys.
\end{abstract}





\section{Introduction}\label{intro}

AGNs are mostly found in X-ray surveys where they represent the majority of
the detected population. However, some AGNs are missed even in the deepest
X-ray observations, e.g. the \chandra\ Deep Surveys,
CDFN~\citep{brandt01,barger02} and CDFS~\citep{giacconi02,rosati02}. The
existence of a large population of AGNs that is still undetected, even in
the deepest X-ray surveys, is suggested by the unresolved X-ray
background~\citep{worsley05} and by the discovery of AGNs not detected at
X-ray energies, but identified at radio and infrared
wavelengths~\citep{polletta06,higdon05,donley05}. Multi-wavelength
observations of AGNs are necessary to make a census of AGNs and to fully
characterize them.

Current facilities and projects permit to study large AGN samples throughout
the entire wavelength spectrum, from radio wavelengths to X-ray energies.
Several wide and deep multi-wavelength surveys are currently being performed
providing enough area, coverage and sensitivity to sample the
multi-wavelength properties of AGNs. A difficult task that multi-wavelength
surveys need to face is the identification of AGN. AGNs usually represent
only a small fraction of sources compared to the population detected at
optical and IR wavelengths and their properties can be elusive. Obscured
AGNs can be elusive because their optical emission resembles that of normal
galaxies and therefore they cannot be identified by the color-selection
techniques used for QSOs (e.g. Schmidt \& Green 1983), or because their
spectra do not show any AGN signature. Their IR emission can be dominated by
PAH features, typical of star-forming galaxies, instead of continuum
emission from hot dust. Radio observations reveal the radio-loud population
of obscured AGNs (e.g. radio galaxies), but this comprises only a small
fraction, $\sim$10\%, of the total AGN population. A hard X-ray selection
fails to detect the luminous narrow-line AGNs in the numbers predicted by
some models of the CXRB.

In order to obtain a complete picture of the AGN population, selection
biases need to be taken into account. The fraction of AGNs missed in
observations in narrow energy windows, e.g. X-rays, infrared, or optical, is
not well known. Moreover, the dispersion in the spectral energy
distributions (SEDs) of all kind of AGNs needs to be well characterized to
evaluate which and what fraction of AGNs are missed in a wavelength window
and detected in another because of the unmatched depths of the observations
at different wavelengths. The SEDs are also important for evolutionary
models~\citep{treister04,silva04,xu03,ballantyne06}, and to estimate the
contribution of accretion to the cosmic backgrounds.

In this work, we characterize the SEDs of a sample of AGNs selected in the
hard X-rays and investigate how they vary as a function of X-ray absorption,
luminosity, redshift, and host galaxy contribution. A selection in the hard
X-rays minimizes the bias against absorbed AGNs compared to samples selected
at soft or broad-band X-ray energies. Our analysis is based on the sources
detected by \xmm\ in a 1 \sqdeg\ region of the XMM-Newton Medium Deep
Survey~\citep[XMDS; ][, hereinafter Paper I]{chiappetti05} in the 
XMM-Large Scale Survey field~\citep[XMM-LSS; ][]{pierre04,pierre07}. This
region has also been observed as part of the \spitzer\ Wide-Area
Infrared Extragalactic Survey (SWIRE)~\citep{lonsdale03} legacy program, of
the VIMOS VLT Deep Survey~\citep[VVDS;][]{mccracken03,lefevre04a}, and of
the Canada-France Hawaii Telescope Legacy ``Deep'' (D1) Survey. The
available multi-wavelength data set, the source catalog, and the X-ray
observations and data are summarized in Section~\ref{dataset} and described
in detail in a companion paper~\citep[][ hereinafter Paper II]{tajer07}.
Here, we present the technique used to classify the SEDs and to estimate
photometric redshifts (Section~\ref{photoz}). We divide the sample in three
main AGN classes based on their SEDs: AGN1, AGN2, and star-forming (SF) like
AGNs. The average properties of the SEDs of the three AGN classes are
described in Section~\ref{agn_classes_seds}. Luminosities and absorption are
estimated and their role in shaping the observed SEDs are discussed in
Section~\ref{lums_abs}. The relationship between the X-ray and the mid-IR
luminosities is investigated in Section~\ref{lmir_lx_ratio}, and the
variations of the mid-IR over X-ray luminosity ratio as a function of X-ray
luminosity and AGN class are discussed in Section~\ref{lmir_lx_ratio}. The
average SEDs of the three classes regrouped by X-ray luminosity are
discussed in Section~\ref{seds_vs_lx}. The analysis of the AGN IR colors
compared with their luminosity, redshift and SED classification is presented
in Section~\ref{agn_colors}. For a sub-set of sources, radio data are also
available. The known radio-IR and radio-X-ray correlations for AGNs and
normal galaxies are compared with our data in Section~\ref{radio}. Our
results are discussed in Section~\ref{discussion}, and summarized in
Section~\ref{summary}.

Throughout the paper, we adopt a flat cosmology with H$_0$ = 71 \kmsMpc,
$\Omega_{M}$=0.27 and $\Omega_{\Lambda}$=0.73~\citep{spergel03}. For
brevity, we refer to the sources in the sample by their XMDS sequence 
number (XMDS seq). The names of the sources, complying the IAU standard,
along with the associated identifiers, are reported in Table~\ref{lum_tab}.
We use the terms red and blue to describe the broad-band SED shape, where
red and blue mean, respectively, increasing and decreasing fluxes, in $\nu
F_{\nu}$, at longer wavelengths. Fluxes refer to observed, not absorption or
k-corrected values, unless specified otherwise. The term absorbed refers to X-ray
sources with effective column densities \nh$\geq$10$^{22}$\cm2, and obscured
to sources with red optical and IR spectra implying dust extinction. The
X-ray luminosity refers to the rest-frame broad band (0.5--10~keV)
luminosity corrected for intrinsic absorption, unless specified otherwise.
We use the terms type 1 and type 2 AGNs for AGNs with broad and narrow
optical emission lines, respectively, in their optical spectra. We introduce
a new nomenclature for the broad-band SED classification: AGN1, AGN2, and
SF. The term AGN1 refers to an SED that is fitted by a type 1 AGN template,
the term AGN2 refers to an SED that is fitted by a type 2 AGNs or composite
template, and the term SF, which stands for star-forming, refers to an SED
that is fitted by a star-forming galaxy template, like spirals and starburst
galaxies. Although the templates are classified on the basis of their
optical spectral properties, we do not expect a perfect match between the
SED classification and the properties of the optical spectra. The SED
classification is more indicative of the main emitting components dominating
at optical and IR wavelengths rather than of their optical spectra.
Moreover, optical spectra can be affected by host galaxy
dilution~\citep{moran02}, obscuration~\citep{barger01,rigby06}, instrumental
and observational limitations, e.g. the lack of emission lines for sources
in certain redshift ranges. 


\section{Sample and multi-wavelength data set}\label{dataset}

The sample studied in this work includes 136 sources detected at $\geq$
3$\sigma$ in the 2--10~keV X-ray band, corresponding to $\simeq$10$^{-14}$
\ergcm2s, in a 1 deg$^2$ region of the XMDS. The XMDS consists of 19 \xmm\
pointings of 20 ksec nominal exposure covering a contiguous area of about
2.6 deg$^2$. The XMDS survey is complete to a 2--10~keV flux of
9$\times$10$^{-14}$\ergcm2s (Paper I). The 1 deg$^2$ area chosen for this
study is centered at $\alpha_{2000}=02^h26^m$, and
$\delta_{2000}=-04^{\circ}30^{\prime}$ (see Paper II for the layout of the
selected field on the entire XMDS area) and benefits from a wealth of
observations from various surveys: \xmm\ data from the XMM-LSS
Survey~\citep{pierre04,pierre07}; optical multi-band broad photometric data
from the VIMOS VLT Deep Survey~\citep[VVDS;][]{mccracken03,lefevre04a}, and
from the Canada-France Hawaii Telescope Legacy ``Wide'' (W1) and ``Deep''
(D1) Surveys (CFHTLS\footnote{http://www.cfht.hawaii.edu/Science/CFHTLS/});
near-infrared broad band photometric data from the UKIRT Infrared Deep Sky
Survey~\citep[UKIDSS;][]{dye06,lawrence06}; infrared data from the \spitzer\
Wide-Area Infrared Extragalactic Survey~\citep[SWIRE;][]{lonsdale03}; and
radio data from the VLA VIMOS Survey~\citep{bondi03,ciliegi05}. Optical
spectroscopic data are also available from the VVDS survey for a small
sub-sample~\citep{lefevre05,gavignaud06}, from observations of a selected
sample of IR-selected AGNs~\citep{lacy06a}, and from the 2dF survey (Garcet
et al., in prep.).

The X-ray observations and data are described in Paper I, and the
multi-wavelength catalog and data set are presented in Paper II.  The X-ray data
cover the energy range from 0.3 to 10~keV, and fluxes are measured in 5
different bands, 0.3--0.5, 0.5--2, 2--4.5, 4.5--10, and 2--10~keV. Column
densities are estimated from the hardness ratio HR derived from the soft
(0.5--2~keV) and hard (2--10~keV) counts. For 55 sources with more than 50 net
counts, the column density is estimated through spectral modeling with
XSPEC~\citep{arnaud96} (for more details see Paper II).

The VVDS data set includes broad-band photometric data in the B, V, R, and I
filters, to an AB limiting magnitude (50\% completeness for point sources)
of 26.5, 26.2, 25.9, and 25.0, respectively~\citep{mccracken03}. Imaging
data in the U band to an AB limiting magnitude of $\sim$25.4 are
also available. The CFHTLS data set includes broad-band photometric data in
the $u^*$, \gp, \rp, \ip, and $z$ filters, to an AB limiting magnitude (50\%
completeness) of \ip=24.5 in the wide survey and \ip=26.1 in the deep survey.

Near-infrared data in the J and K bands are available for a small fraction
($\sim$4\%) of the area down to an AB limiting magnitude (50\% completeness)
of 24.2 and 23.9. Additional near-infrared data from the UKIDSS project are
also available. The UKIDSS observations cover about 54\% of the XMDS area
and include J and K band data to a limiting magnitude of 22.3 and 20.8,
respectively.

The \spitzer\ infrared data include broad-band photometric data from the
four IRAC bands at 3.6, 4.5, 5.8, and 8.0~$\mu$m to a 5$\sigma$ limit of 4.3,
8.3, 58.5, and 65.7~$\mu$Jy, respectively, and MIPS data at 24~$\mu$m to a 5$\sigma$
limit of 241~$\mu$Jy. The SWIRE observations cover about 90\% of the area and
are available for 122 sources, 91\% of the entire sample.

The radio data are from the VLA VIMOS Survey which reaches a 5$\sigma$
depth of 80~$\mu$Jy at 1.4~GHz~\citep{bondi03,ciliegi05}.

The associations between the X-ray source list and the multi-wavelength
catalogs are described in detail in Papers I and II. The selected
multi-wavelength catalog contains 136 sources. Two sources are galaxy
clusters (see Paper II) and are not included in the following analysis. Of
the remaining 134 sources, 130 have optical counterparts, unique for 121 and
multiple candidates for the other remaining 9 sources. Out of four sources
without optical counterpart, one is close to a bright star and the
photometry could not be extracted and the remaining 3 sources are optically blank
fields. All of the 122 sources that fall into the SWIRE area are detected in
at least two \spitzer\ bands, and 87 are also detected at 24~$\mu$m. UKIDSS
data are available for 66 sources, of which 55 are also SWIRE sources and 64
have an optical counterpart. Radio data are available for 32 sources, of
which 30 are also SWIRE sources and 30 have an optical counterpart.

\section{Template fitting \& photometric redshift technique}\label{photoz}

In order to classify the spectral energy distributions (SEDs) and estimate
photometric redshifts of the sources in the sample, optical and IR data are
combined and fitted with a library of galaxy and AGN templates. 
 \begin{figure*}
  \epsscale{2.0}
   \plotone{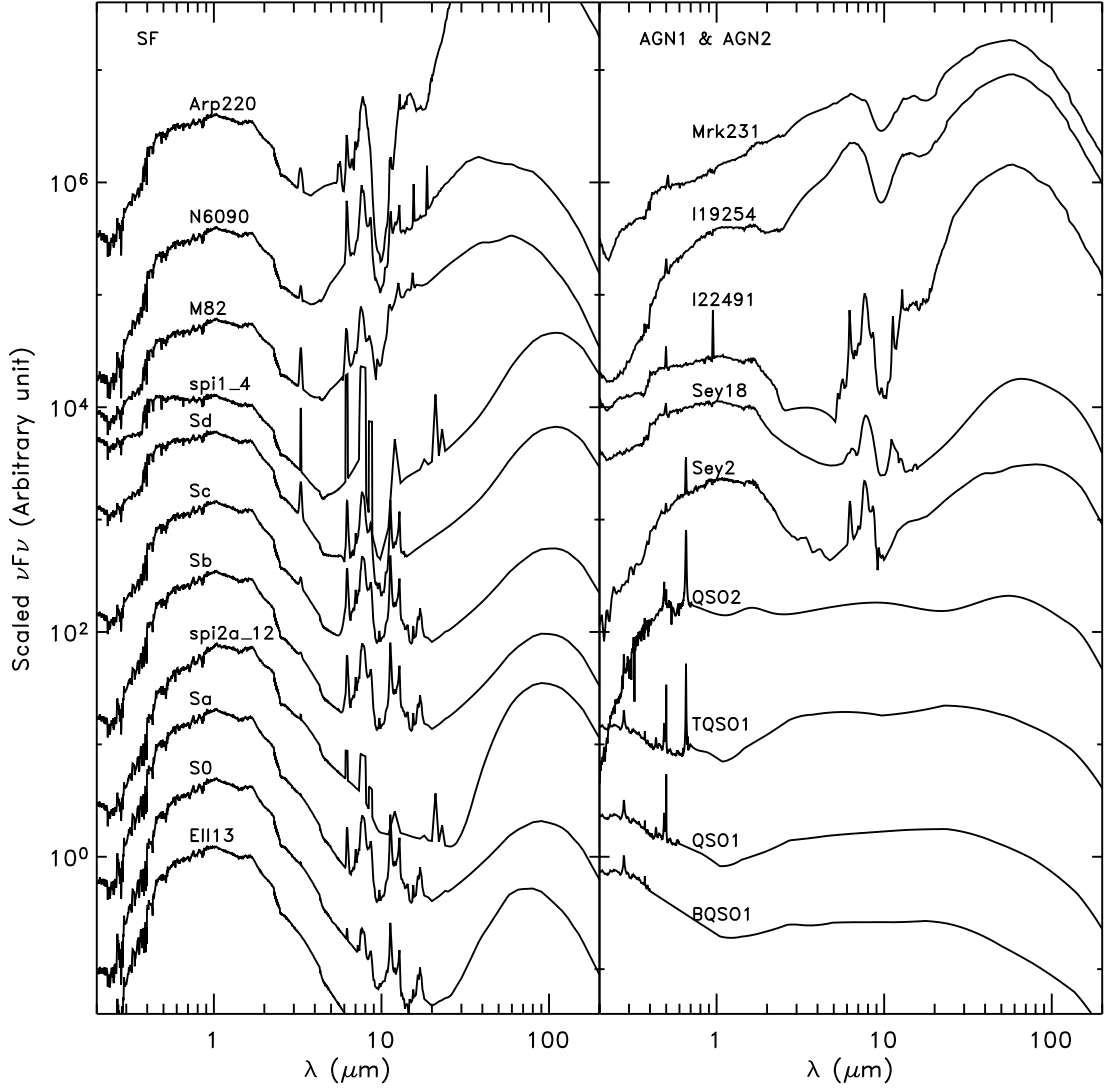}
      \caption{SEDs in $\nu F_{\nu}$ $versus$ $\lambda$ of the template
               library used in this work. Elliptical, spiral and starburst
               galaxy templates are shown in the left panel and AGN
               templates in the right panel, the type 2 AGNs are the top 6
               and the type 1 AGNs are the bottom 3. The names of the
               templates are annotated.}
         \label{templates}
   \end{figure*}
The SEDs are fitted using the Hyper-z code~\citep{bolzonella00}. 
Hyper-z has been already successfully applied to various data sets and
offers the possibility of using any template library and filters set. 
Hyper-z finds the best-fit by minimizing the $\chi^2$ derived from the
comparison of the observed SED and expected SEDs, at various redshifts,
derived from the templates using the same photometric system. The effects of
dust extinction are taken into account by reddening the reference templates
according to a selected reddening law. We use the prescription for
extinction derived from high-redshift starbursts by~\citet{calzetti00}. In
order to limit degeneracies in the best-fit solutions we limit the
extinction \av\ to be less than 0.55 mag and include templates of highly
extincted objects to fit heavily obscured sources.

The library contains 20 templates including 1 elliptical, 7 spirals, 3
starbursts, 6 AGNs, and 3 composite (starburst+AGN) templates covering the
wavelength range between 1000\AA\ and 1000\micron. The elliptical, spiral
and starburst templates were generated with the GRASIL
code~\citep{silva98}. The 7 spirals range from early to late types (S0-Sd),
the starburst templates correspond to the SEDs of NGC 6090, M 82 and Arp
220.  In all of the spirals and starburst templates the spectral region
between 5 and 12~$\mu$m, where many broad emission and absorption features
are observed, was replaced using observed IR spectra from the PHT-S
spectrometer on the {\em Infrared space Observatory} and from IRS on
\spitzer.

Templates of moderately luminous AGN, representing Seyfert 1.8 and Seyfert 2
galaxies, were obtained by combining models, broad-band photometric data
(NED), and ISO-PHT-S spectra (Schulz, private communication) of a random
sample of 28 Seyfert galaxies. The other four AGN templates include three
templates representing optically-selected QSOs with different values of
infrared/optical flux ratios (QSO1, TQSO1, and BQSO1) and one type 2 QSO
(QSO2). The QSO1 templates are derived by combining the SDSS quasar
composite spectrum and rest-frame infrared data of a sample of 35 SDSS/SWIRE
quasars~\citep{hatziminaoglou05a}. After normalizing each SED in the
optical, we derived three templates with the same optical spectrum but three
different IR SEDs. The QSO1 IR template was obtained from the average fluxes
of all the measurements regrouped in wavelength bins, the TQSO1 IR template
was obtained from the highest 25\% measurements per bin, and the BQSO1 IR
template was obtained from the lowest 25\% measurements per bin. The type 2
QSO template (QSO2) was obtained by combining the observed optical/near-IR
spectrum of the red quasar FIRST J013435.7$-$093102~\citep{gregg02} and the
rest-frame IR data from the quasars in the Palomar-Green sample with
consistent optical SEDs~\citep{polletta06}.

The composite (AGN+SB) templates are empirical templates created to fit the
SEDs of the following objects: the heavily obscured BAL QSO Mrk
231~\citep{berta05}, the Seyfert 2 galaxy IRAS 19254$-$7245
South~\citep[I19254;][]{berta03}, and the Seyfert 2 galaxy IRAS
22491$-$1808~\citep[I22491;][]{berta05}. All these objects contain a
powerful starburst component, mainly responsible for their large infrared luminosities
($>$10$^{12}$\lsun), and an AGN component that contributes to the mid-IR
luminosities.

The full library of galaxy and AGN templates is shown in
Figure~\ref{templates}. With respect to existing template libraries derived
from empirical SEDs~\citep{coleman80} or from models~\citep{bruzual03,
fioc97, silva98, devriendt99}, this library has a broader wavelength
coverage and variety of spectral types. Examples of application of this
library to various types of SWIRE sources can be found
in~\citet{lonsdale04,franceschini05,hatziminaoglou05a,jarrett06,polletta06,weedman06}.

Photometric redshift techniques have been applied to optical or optical and
near-infrared data sets providing reliable photometric
redshifts~\citep{rowan-robinson05,ilbert06,brodwin06}. However, with a
limited wavelength coverage, some spectral types can be degenerate, e.g.,
obscured starbursts and ellipticals. In those cases, the inclusion of mid-IR
data can break the degeneracy. However, the inclusion of mid-IR data in
photometric redshift techniques can degrade the photometric redshift
estimates because the templates represent only a limited range of optical/IR
ratios. We compared both approaches, fitting optical+near-IR and
optical+near-IR+mid-IR data, and concluded that the latter method gives 
significantly better results both in the photometric redshift estimates and
in spectral type classification.

We fit the SEDs using the data from the optical to 24~$\mu$m. Mid-IR
(5.8--24~$\mu$m) data are available for 107 out of the 134 sources in the
sample. The best solution corresponds to the solution with minimum $\chi^2$
among all of the solutions with $z$=0--4 and B-band absolute magnitude
within a pre-defined range. The allowed range of B-band absolute magnitude,
M$_\mathrm{B}$, varies with redshift and is different for normal galaxies
and for AGN templates. M$_\mathrm{B}$ must be greater than
M$_\mathrm{B}^{min}=-6-5\times$Log(d$_L$), where d$_L$ is the luminosity
distance in Mpc, and lower than M$_\mathrm{B}^{max}=-0.7-5\times$Log(d$_L$),
but always within the range
$-$23.7, $-$17 for normal galaxy templates and greater than
M$_\mathrm{B}^{min}=-6.5-5\times$Log(d$_L$) and lower than
M$_\mathrm{B}^{max}=-5\times$Log(d$_L$), but always within the range
$-$28.8, $-$19 for AGN templates. The best-fit solution is the one with the
lowest $\chi^2$ among all of the solutions that satisfy the
absolute-magnitude criterion. If there are no solutions in the defined
absolute magnitude range, the criterion is modified by extending the limits
M$_\mathrm{B}^{min}$ and M$_\mathrm{B}^{max}$ by steps of 0.5 magnitudes,
but always within the maximum allowed range, $-$23.7, $-$17 for normal
galaxy solutions and $-$28.8, $-$19 for AGN solutions. The criterion was
relaxed in 7 cases (see M flag in Table~\ref{lum_tab}).

\begin{figure}
   \epsscale{1.0}
   \plotone{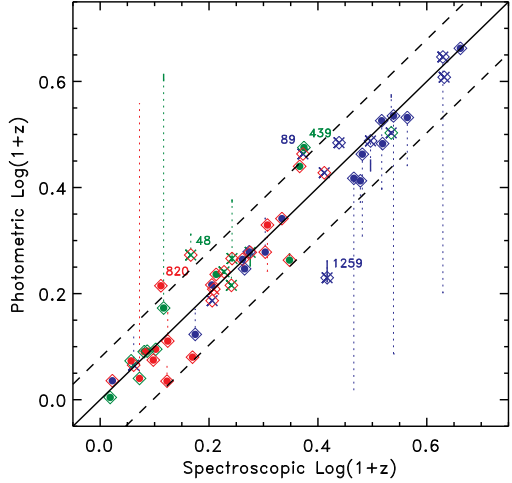}
      \caption{Comparison between photometric and spectroscopic redshifts in
       Log(1+$z$) of the 49 sources with available spectroscopic redshifts
       (open diamonds). Dashed lines represent 20\% agreement in
       (1+$z$). The 5 catastrophic outliers are labeled with their XMDS seq:
       48, 89, 439, 820, and 1259. Full circles (33 sources)
       represent the sources in the selected sample presented in this work. 
       Crosses (16 sources) represent the additional sources in the XMDS
       sample with spectroscopic redshifts. Circles and plus signs are
       color-coded according to the SED class (blue: AGN1; red: AGN2; green:
       SF).  A diamond is overplotted on each source and is color-coded
       according to the spectroscopic classification (blue: type 1 AGN; red:
       type 2 AGN; green: normal galaxy). The vertical dotted lines show the range
       of photometric redshifts of all acceptable solutions (note that the
       acceptable solutions can be at the extremes of the bar and 
       not cover a continuous range in redshifts.}
         \label{zphot_zspe}
   \end{figure}

In addition to the best solution, we consider up to two secondary
solutions within the defined absolute magnitude limits. In most of the cases
there are three solutions per source (the best and two secondary solutions),
but in some cases, only one solution (or one minimum in the $\chi^2$
distribution), is found within the defined absolute magnitude limits, thus
for each source there might be from one to three solutions. We will refer to
these as the acceptable solutions. The comparison between the spectroscopic
and final photometric redshifts is shown in Figure~\ref{zphot_zspe}.

The reliability and accuracy of the photometric redshifts are measured via
the fractional error $\Delta z$, the systematic mean error 
$\overline{\Delta z}$, the 1$\sigma$ dispersion $\sigma_z$, and the
rate of catastrophic outliers, defined as the fraction of sources with
$\left| \Delta z\right| >$ 0.2. $\Delta z$ is defined as:
\begin{equation}
\Delta z = \left(\frac{z_{phot}-z_{spec}}{1+z_{spec}}\right)
\end{equation}
and
\begin{equation}
\sigma_z^2 = \sum \left(\frac{z_{phot}-z_{spec}}{1+z_{spec}}\right)^2/N
\end{equation}
with N being the number of sources with spectroscopic redshifts. Since
there are only 33 sources in the selected sample with a spectroscopic
redshift measurement (see Table~\ref{lum_tab}), we also included in the
analysis of the photometric redshifts, all of the sources in the XMDS
4$\sigma$ VVDS sample (Paper I) with a spectroscopic redshift measurement,
for a total of 49 sources.  The systematic mean error, $\overline{\Delta
z}$, is $-$0.001 ($-$0.008), the $rms$, $\sigma_z$, is 0.12 (0.11), and the
outlier fraction is 10\% (6\%) for the 49 sources with spectroscopic
redshifts. The values in parenthesis correspond to the results obtained
using only the 33 sources with spectroscopic redshifts in the hard X-ray
selected sample. Although these results are not as satisfactory as those
achieved for galaxy populations, even when few broad-band photometric data
are available~\citep{babbedge04}, they are better than what has been
previously obtained for AGN samples. The fraction of outliers for AGNs is
indeed usually higher than 25\%~\citep{kitsionas05,babbedge04}. The achieved
accuracy does not allow us to perform detailed analysis on single sources.
However, it is adequate for a statistical analysis of the population as
presented in this work. Four (XMDS seq: 48, 89, 439, and 820) of the 5
outliers have $\left| \Delta z\right| <$0.28 and two of them (XMDS seq: 439, and 820) are in the
hard X-ray sample. The only outlier with $\left| \Delta z\right| >$0.28
is source XMDS seq 1259, which is not included in the selected sample.

 \begin{figure}
   \epsscale{1.0}
   \plotone{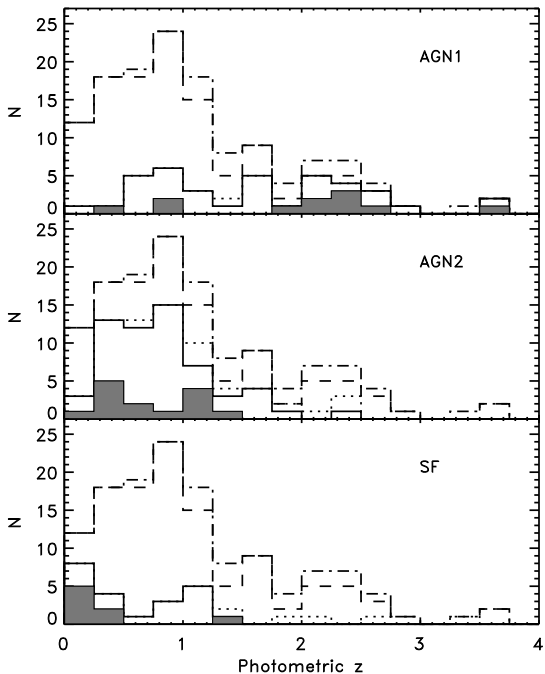}
      \caption{Redshift distribution of the entire sample (samples A+B), 134
       sources (dash-dotted line), and of sample A only, 119 sources (dashed
       line) (see Section~\ref{agn_classes_seds}). Each panel shows the
       redshift distributions of the sources classified as AGN1 (top panel),
       AGN2 (middle panel), and SF (bottom panel). The dotted lines
       correspond to the sources in each class in the entire sample, and the
       solid lines only to those in sample A. The shaded area represents the
       distribution of spectroscopic redshifts in each class.}
         \label{histo_z}
   \end{figure}

The redshift distributions of the whole sample,
based on the photometric redshifts is shown with a dot-dashed line
in Figure~\ref{histo_z}. The distribution peaks at $z\simeq$0.9, consistent
with the redshift distribution of other X-ray selected
samples~\citep{hasinger02,barger02,szokoly04,eckart06}. About 54\% (73
out of 134) of the sources are at redshift below 1 and 90\% are at redshift
below 2.3. In the Figure, the redshift distribution of the total sample is
compared with that of AGN1, AGN2, and SF sources (see Section~\ref{seds}).

\subsection{Spectral energy distributions}\label{seds}

 \begin{figure*}
  \epsscale{2.0}
   \plotone{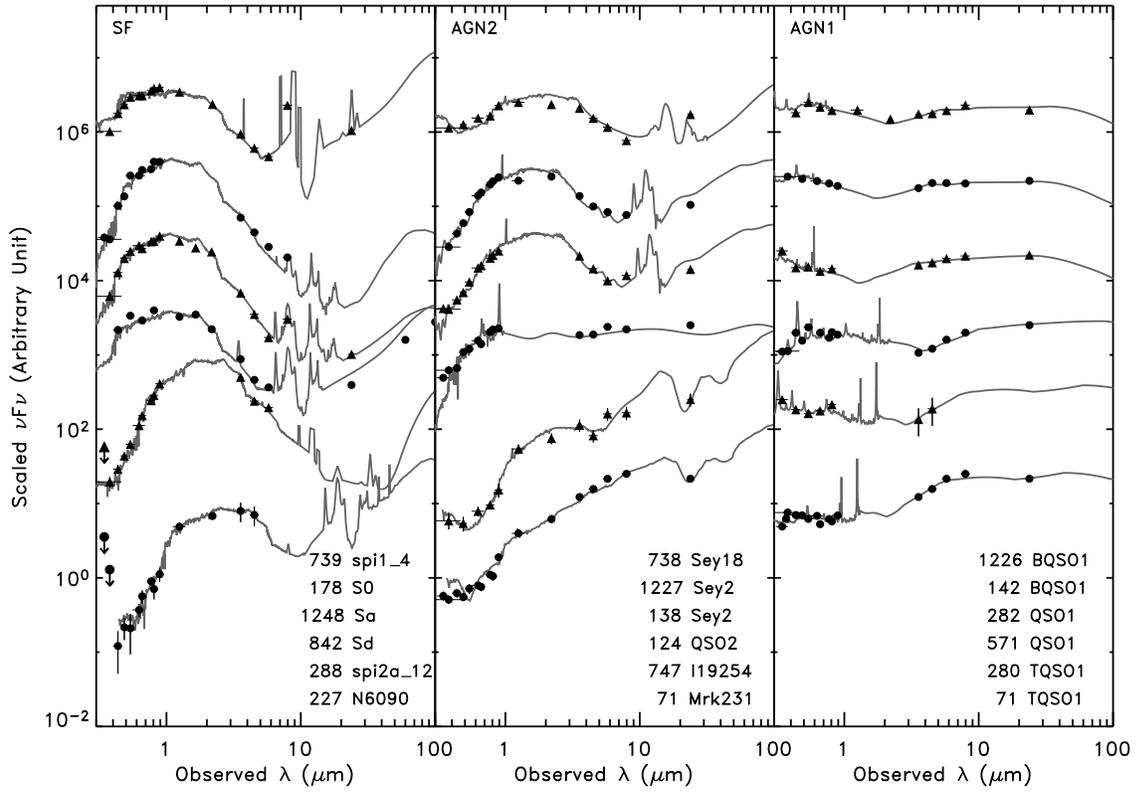}
      \caption{Scaled SEDs (full black circles and triangles) and best-fit templates
               (gray solid curves) in $F_{\nu}$ $versus$
               $\lambda$ of 18 sources: 6 classified as SF (left panel), AGN2
               (middle panel), and AGN1 (right panel).  Downward pointing
               arrows correspond to 5$\sigma$ upper limits. The source
               sequence number and best-fit template names are listed in the
               same order as the SEDs are plotted.}
         \label{sed_fits}
   \end{figure*}

According to the template corresponding to the best-fit solution selected by
the procedure described above, each source is classified as one of the
following broad classes, AGN1, AGN2, or SF. Examples of fits for the three
classes are shown in Figure~\ref{sed_fits}. The AGN1 class includes all of
the sources best-fitted with a QSO1 template (bottom three on the right
panel of Figure~\ref{templates}, see also right panel of
Figure~\ref{sed_fits}). They are characterized by blue optical SEDs
($\lambda F_{\lambda} \propto \lambda^{-0.9,-0.4}$) and red IR SEDs
($\lambda F_{\lambda} \propto \lambda^{0.2,0.4}$). The change in slope, from
blue to red, occurs at about 1~$\mu$m. Emission from the accretion disk
dominates at shorter wavelengths and from AGN-heated dust at longer
wavelength. The AGN2 class includes sources best-fitted with the Seyfert
templates, the three composite AGN+SB templates, or the QSO2 template (top
six on the right panel of Figure~\ref{templates}, and see also middle panel
of Figure~\ref{sed_fits}). The AGN2 templates include a wide variety of SED
shapes. They are mostly red at optical and near-IR wavelengths, but in some
cases an upturn (blue) at short wavelengths (far-ultraviolet) is present.
Their IR SEDs range from red power-law like SEDs, similar to the AGN1 or
redder, to composite SEDs where emission features associated with
star-formation are present in addition to continuum emission in the near-IR.
The SF class includes sources fitted with a spiral or a starburst template
(left panel of Figures~\ref{templates} and~\ref{sed_fits}). SF templates are
all red in the optical and blue in the near-IR, consistent with emission
from stellar populations. At mid-IR wavelengths, the SED is dominated
by emission features and a warm dust continuum associated with star-forming
regions. Photometric and spectroscopic redshifts, best-fit template and
classification are reported in Table~\ref{lum_tab}. 

Although a best-fit template, and thus an SED classification, is available
for all sources in the sample, in the following analysis we will discuss the
results only for sources with a reliable fit and simply report the derived
values for the rest of the sample for completeness. We define a fit reliable
if at least six data points could be fitted. Six sources (XMDS seq: 139,
199, 233, 403, 840, 1219) have been observed only in the optical or in the
IR and have less than six data points in total. Because of the limited
wavelength coverage, they are not included in the following analysis. Nine
sources with ambiguous counterpart (XMDS seq: 91, 111, 120, 227, 255, 449,
470, 747, 844) are also excluded in the following analysis. We will
refer to the 15 sources as sample B. In Table~\ref{lum_tab} we report all
best-fit solutions for the sources with ambiguous counterpart in order of
increasing distance of the optical counterpart, but in the following figures
we will show only the best-fit solution corresponding to the closest
counterpart. The sample that will be analyzed in the following sections
contains the remaining 119 sources (the total 134 minus the 15 in sample B;
hereinafter sample A), of which 38 are classified as AGN1s (32\% of the
sample), 59 (50\%) as AGN2s and the remaining 22 (18\%) as SFs.

The reliability of this classification depends on the number of available
optical and IR data points and on the ability of identifying the signatures
that make the SEDs of the various classes unique. In the case of AGN1s, the
blue optical SED and red IR SED provide two clear signatures that make their
classification highly reliable (see right panel in Figure~\ref{sed_fits}).
The distinction between AGN2s and SFs is less obvious than the
identification of AGN1 since the only difference between their SEDs is in
the shape of the near- and mid-IR SED (see left and middle panels in
Figure~\ref{sed_fits}). When only few IR data points are available or the
data do not sample the region where the AGN signature is visible (i.g.
3--5~$\mu$m rest-frame), the separation between the two classes is more
uncertain. In order to estimate the uncertainty of our classification we
counted all sources with an acceptable solution in a different class (see
Section~\ref{photoz} for a description of the acceptable solutions). For
each class, we counted the number of sources with solutions in other
classes, and the number of sources in other classes with a secondary
solution in the specific class. For the AGN1 class, we estimate that 2
sources might belong to a different class, AGN2 or SF, 4 AGN2s have an AGN1
secondary solution, and 2 SFs have an AGN1 secondary solution. Thus, there
are 36 sources with a secure AGN1 fit, and 44 sources with a possible AGN1
fit, which implies an AGN1 fraction of 32$^{+5}_{-2}$\%. Analogously, for
the AGN2 class, we estimate that 13 sources might belong to a different
class, AGN1 or SF, 1 AGN1 has an AGN2 solution, and 5 SFs have an AGN2
solution. Thus, there are 46 sources with a secure AGN2 fit, and 65 sources
with a possible AGN2 fit, which implies an AGN2 fraction of
50$^{+5}_{-11}$\%. For the SF class, we estimate that 7 sources might belong
to a different class, AGN1 or AGN2, 1 AGN1 has an SF solution, and 9 AGN2s
have an SF solution. Thus, there are 13 sources with a secure SF fit, and 30
sources with a possible SF fit, which implies an SF fraction of
18$^{+7}_{-7}$\%. Note that, by definition, the given uncertainties on the
class fractions correspond to lower and upper limits to the fraction of
sources in each class.

The reliability and completeness of our classification can be estimated
using the spectroscopic classification, however we do not expect an exact
correspondence between our SED-based classes and the spectroscopic classes.
A spectroscopic classification is available for 49 sources, including
sources that are not in the hard X-ray samples. Of these, 26 are
spectroscopically classified as type 1 AGNs, 12 are spectroscopically
classified as type 2 AGNs, and 11 as star-forming galaxies. Among the 26
type 1 AGNs, 16 are classified AGN1s, 9 AGN2s, and 1 SF, and all the AGN1s
with a spectroscopic classification (16) are type 1 AGNs. Thus, the SED
classification is 62\% complete (16 AGN1s out of 26 type 1 AGNs), and 100\%
reliable (all AGN1s are type 1 AGNs). Among the 12 sources spectroscopically
classified as type 2 AGNs, 7 are classified as AGN2s, and the remainings as
SFs. The AGN2 class (24 sources) includes 9 type 1 AGNs, 7 type 2 AGNs, and
8 SFs. Thus, the completeness of the AGN2 sample is 58\% (7 AGN2 out of 12
type 2 AGNs), and the reliability is 29\% (7 out of 24 AGN2s are type 2
AGNs). There are 11 sources with a spectroscopic classification consistent
with being normal galaxies. Eight of them are classified as AGN2s, and 5 as
SF. Among the 9 sources classified as SF, 1 is a type 1 AGN, 5 are type 2
AGNs, and 3 are star-forming galaxies according to the spectroscopic
classification. The completeness of the SF sample is thus 27\% (3 out of 11
star-forming galaxies are classified SF), and the reliability is 33\% (3 out
of 9 SF sources are star-forming galaxies). In summary, the SED and the
spectroscopic classification agree in 26 of the 49 sources with
spectroscopic classification (53\%).

We can also compare the fraction of AGNs that are identified as AGNs by the
SED classification (AGN1 or AGN2) and by the spectroscopic classification
(type 1 and type 2 AGNs). Note that all sources are AGNs as probed by their
X-ray properties. Out of the 49 AGNs with spectroscopic classification, 38
(78\%) are spectroscopically identified as AGNs (26 type 1 and 12 type 2),
and 40 (82\%) are classified AGNs by the SED fitting technique (25 AGN1 and
7 AGN2). The latter fraction is consistent with the 83\% derived from the
entire X-ray sample.

\subsection{Redshift-dependent derived quantities: luminosities and
absorption}\label{lum_z}

Using the available redshifts, spectroscopic if available, photometric
otherwise, and the best-fit templates, we derive the luminosities in the
broad and hard X-ray energy bands (0.5--10~keV and 2--10~keV rest-frame), in
the mid-IR (3--20~$\mu$m rest-frame), and in the radio at 1.4~GHz
rest-frame. The X-ray luminosities, hereinafter $L_{0.5-10~keV}$ and
$L_{2-10~keV}$, are derived assuming a power-law spectrum
(F(E)$\propto$$E^{-(\Gamma-1)}$) with photon index $\Gamma$=2.0. Effective
column densities are derived from the observed column densities after
applying the redshift correction (see Paper II for details). In case the
observed column density is consistent with the Galactic value, we derive an
upper limit to the effective column by assuming the Galactic value and the
redshift of the source (see also Section~\ref{lums_abs}). The observed
column densities are used to correct the X-ray luminosity for absorption,
hereinafter $L_{0.5-10~keV}^{corr}$. The uncertainty on the X-ray
luminosities are estimated by propagation of the uncertainty on the flux and
on the column density.

The mid-IR luminosities, $L_{3-20 \mu m}$, are estimated by integrating the
best-fit template in the rest-frame 3--20~$\mu$m wavelength range. To
estimate the uncertainty on the mid-IR luminosities we first fit the IR data
($\lambda_{obs}>3.6\mu$m) of each source with all of the templates, derive
the $\chi^2$ and then use the range of luminosities obtained from all
acceptable solutions.

The radio luminosities, $L_{1.4~GHz}$, are derived assuming a power-law
spectrum (F$_{\nu}\propto \nu^{-\alpha_R}$) with spectral index
$\alpha_R$=0.8~\citep{condon88}. The uncertainty on the radio luminosity is
derived from the radio flux uncertainty.

The X-ray, mid-IR and radio luminosities are listed in Table~\ref{lum_tab}.

\subsection{Star-forming galaxies}\label{sfg}

In the following analysis, we will include only sources that contain an AGN.
In order to identify sources whose X-ray emission is consistent with
emission from a star-forming galaxy, e.g., from X-ray binaries, we require a
broad X-ray luminosity below 10$^{42}$\ergs, and a soft X-ray spectrum.
Star-forming galaxies are typically characterized by luminosities below
10$^{42}$\ergs and show steep X-ray spectral
slopes~\citep{kim92a,kim92b,colbert04,persic04}. This criterion might
misidentify some normal star-forming galaxies with a large population of
high-mass X-ray binaries (HMXBs) that produces a flat X-ray
spectrum~\citep{colbert04},~\citep[see also Figure 2 in][]{alexander05a},
and might also misidentify low-luminosity AGNs whose X-ray luminosity can be
lower than 10$^{42}$\ergs.

There are 3 sources with an absorption-corrected luminosity below
10$^{42}$\ergs, sources XMDS seq 178, 842 (aka Arp 54), and 1248. Sources
XMDS seq 178 and 1248 are characterized by low column densities,
22$\times$10$^{20}$\cm2 and 2.6$\times$10$^{20}$\cm2 (consistent with
the Galactic value), and hardness ratios, $-$0.43, and $-$0.54,
respectively. However, source XMDS seq 842 ($z_{spec}$=0.043) has a hardness
ratio of 0.16 which implies a column density of 1.78$\times$10$^{22}$ \cm2. 
Because of its hard X-ray spectrum, source XMDS seq 842 is included in the
AGN sample that will be investigated in the following sections. The
classification of source 178 is also dubious because its optical to X-ray
flux ratio ($Log(F_{0.5-10~keV}/F_R)\simeq -0.49$) is consistent with those
typical of unobscured AGNs~\citep[$-$1$<Log(F_{0.5-10~keV}/F_R)<1$;][]{akiyama03}. 
Indeed in Paper II, where star-forming galaxies are identified based on the
optical to X-ray flux ratio, XMDS seq 178 is considered an AGN and XMDS seq
842 is considered a star-forming galaxy. Here, we prefer to adopt a
criterion which is based only on intrinsic X-ray properties (luminosity and
\nh) rather than flux ratios that are redshift-dependent. Thus, we consider
XMDS seq 178 and 1248 normal star-forming galaxies, and XMDS
ID 842 an AGN. Sources XMDS seq 178 and 1248 will not be taken into account
in the following analysis which will focus only on the AGN population (132
sources in total, and 117 in sample A).

The fractions of AGNs in the three SED classes thus become 33${+5}_{-1}$\%
AGN1s, 50${+6}_{-11}$\% AGN2s, and 17${+9}_{-6}$\% SFs.

\section{Average SEDs of the three AGN SED classes}\label{agn_classes_seds}

  \begin{figure}
   \epsscale{1.0}
    \plotone{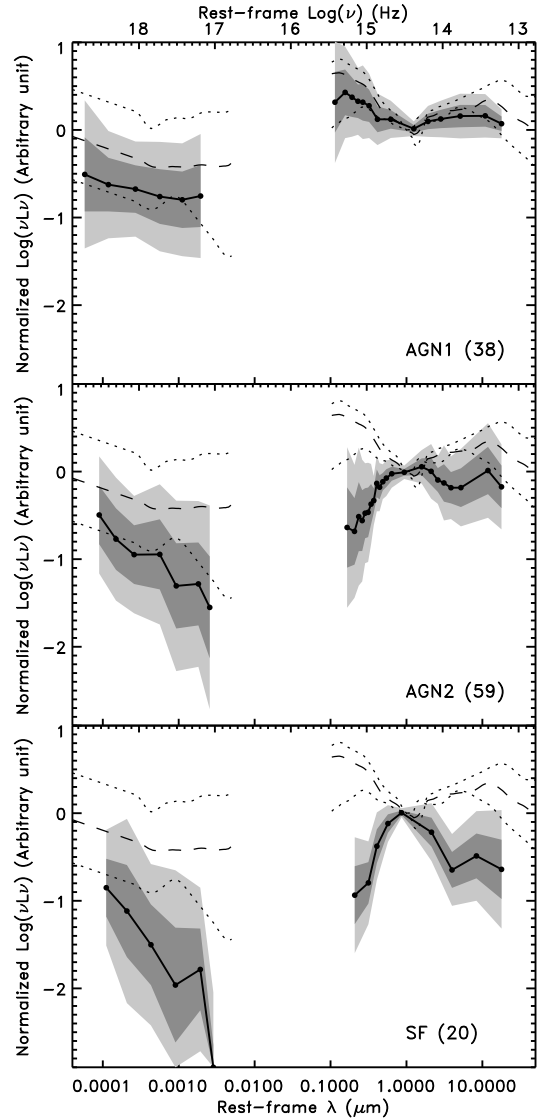}
    \caption{Normalized at 1~$\mu$m average rest-frame X-ray--mid-IR SEDs in Log($\nu
     L\nu$) $vs$ Log($\lambda$) (full circles connected by straight lines)
     of AGNs classified as AGN1 (38 sources) (top panel), AGN2 (59 sources)
     (middle panel) and SF (20 sources) (bottom panel). The light and dark shaded
     areas correspond to 2 and 1$\sigma$ dispersion, respectively.
     Note that the X-ray luminosities are not corrected for absorption.
     The black thin curves represent the median (dashed curve) $\pm$90\%
     dispersion (dotted curves) QSO template by~\citet{elvis94a}.}
              \label{rf_seds}
    \end{figure}
After deriving the luminosity SEDs in $\nu$L$\nu$ for each source assuming
the source photometric redshift, or spectroscopic redshift if available,
average normalized rest-frame SEDs, from the X-ray to the mid-IR, were
obtained from all AGNs in the A sample (117 sources) divided by class (SF,
AGN2 and AGN1). The average SEDs are shown in Figure~\ref{rf_seds}.
Similarly to the median QSO template in~\citet{elvis94a}, we normalize the
SEDs at 1~$\mu$m rest-frame assuming the best-fit template for each source. 
This choice of wavelength is well suited to investigate separately the SED
dispersion of the main emitting components in AGNs, the accretion disk and
the stars in the optical and the dust in the infrared. The average SEDs are
derived from the normalized rest-frame SEDs by taking the mean of the
logarithm of the wavelength and the weighed mean of the luminosity in
Log($\nu$L$\nu$) in 0.3 wide bins. Smaller bins were chosen if more than 40
data points were available or wider bins to include at least 10 points. The
weights are based on the luminosity uncertainties. Upper limits in the
optical and infrared are taken into account by assuming the upper limit to
the luminosity and a logarithmic uncertainty equal to twice the maximum
logarithmic uncertainty of each source. Upper limits to the X-ray
luminosities are taken into account by assuming as luminosity the 1$\sigma$
luminosity value and a logarithmic uncertainty equal to twice the maximum
logarithmic uncertainty observed in the X-rays for that source. The average
SF, AGN2, and AGN1 SEDs contain respectively 15, 29, and 20 bins, and are
obtained by averaging the rest-frame SEDs of 20 SFs, 59 AGN2s, and 38 AGN1s.
The average SEDs are shown in Figure~\ref{rf_seds} as thick black curves, 1
and 2$\sigma$ dispersion are shown as grey shaded areas. Since the redshift
distribution of the sample is broad, up to $z\geq$3.5, and the majority of
the sample has
$z\sim$1, the observed X-ray data sample rest-frame energies up to 45 keV
for the highest-$z$ sources, and up to 20 keV for the majority of the
sample. Thus, the average SEDs sample rest-frame energies up to 20 keV.

The global SEDs, from X-ray to infrared, show a wide dispersion in each
class. Since they are normalized at 1~$\mu$m, this dispersion is more
pronounced at X-ray wavelengths. The average SEDs of the three classes show
some clear differences.  The average SEDs become increasingly blue in the
optical-NIR ($\lambda^{rest}\leq 1\mu$m), red in the IR
($\lambda^{rest}\simeq 1-10\mu$m), and soft in the X-rays in the sequence
SF$\rightarrow$AGN2$\rightarrow$AGN1 (from bottom to top in
Figure~\ref{rf_seds}). The average SEDs are compared with the median QSO
template of optically-selected quasars~\citep{elvis94a} in
Figure~\ref{rf_seds}. The average AGN1 SED is redder in the optical than
Elvis's template, with only the bluest sources being consistent with Elvis's
template. The redder optical SED of the AGN1s compared to Elvis's template
suggests that these sources might include some reddened type 1
AGNs~\citep{wilkes02,gregg02,glikman04,urrutia05,wilkes05}. Indeed, the
observed range of extinction for AGN1s in V-band is 0.0--0.55, and the
median value is 0.40 (see Table~\ref{lum_tab}). The SF and AGN2 optical SEDs
are almost identical and significantly redder than the AGN1 optical SEDs. 
The observed different shape suggests that the origin of the optical
emission is different in AGN1 and in AGN2, and SF sources, rather than the
effect of reddening on an AGN spectrum. The AGN2 and SF optical emission is
likely dominated by the host galaxy emission in most of the cases.

The near- and mid-IR ($\lambda\simeq$1--20 ~$\mu$m) average SEDs of
AGN1s and AGN2s are consistent with each other at 1$\sigma$ level, and both
are redder than the SF average SED.  The mid-IR SED of AGN1 and AGN2 are
consistent with being dominated by AGN-heated hot dust emission, while, in
SFs, the host galaxy dominates. The mid-IR SED of Elvis's template is
consistent with the average AGN1 mid-IR SED, and redder than the average
AGN2 mid-IR SED.

In the X-rays, we show X-ray luminosities with no correction for
intrinsic or Galactic absorption. AGN1s show on average softer X-ray spectra
than SFs and AGN2s, although there is a significant dispersion.  The
observed average spectra are consistent with increasing absorption in AGN2
compared to AGN1, and in SF compared to AGN2.

Compared to Elvis's template, the normalized X-ray luminosity
(L$_X$/L(1$\mu$m)) of AGN1s is more than two times lower. This difference might be
due to a well known bias in Elvis's QSO sample which favors sources with
strong X-ray-to-optical luminosity ratios and detections in the soft
X-rays~\citep{wilkes87,elvis94a}.

The observed differences in the optical and IR SEDs of the three classes are
mainly a consequence of the spectral classification. However, the SED
classification is independent of the X-ray properties, therefore the
differences in the X-ray average spectra indicate that harder X-ray spectra
are observed in more obscured AGNs.

\section{Luminosities and X-ray absorption $versus$ AGN SED class}\label{lums_abs}

   \begin{figure}
    \epsscale{1.0}
    \plotone{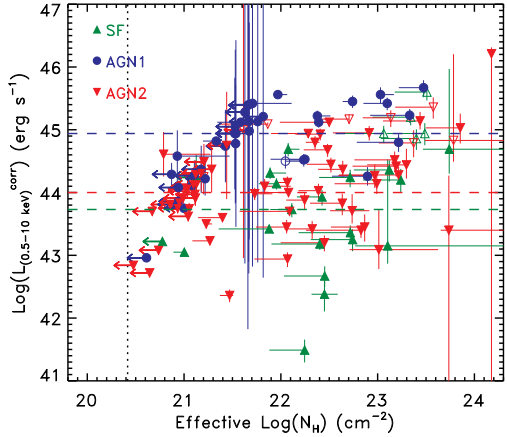}
    \caption{Rest-frame X-ray luminosities corrected for absorption as a
             function of the effective column density measured in the X-ray
             for the whole sample (open and full circles). Full symbols
             represent sources in sample A and open symbols to sources in
             sample B (see Section~\ref{agn_classes_seds}). Different
             symbols and colors correspond to the 3 SED classes, AGN1: blue
             circles, AGN2s: red reversed triangles, and SFs: green
             triangles. The dashed lines represent median values of the
             absorption-corrected X-ray luminosities for the 3 classes.}
             \label{lbx_nhe}
   \end{figure}
   \begin{figure}
    \epsscale{1.0}
   \plotone{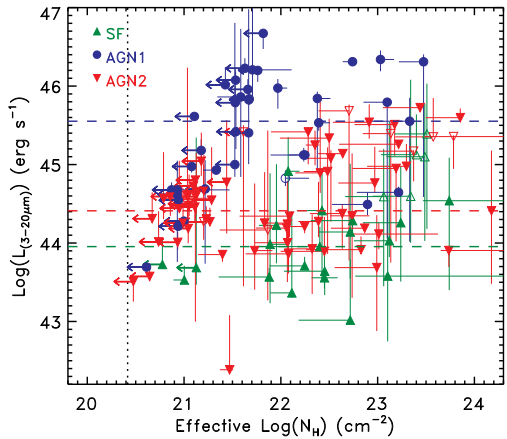}
      \caption{Comparison of mid-IR luminosity with the effective column
      density measured in the X-rays. Symbols as in
      Figure~\ref{lbx_nhe}.}
         \label{lmir_nhe}
   \end{figure}

The absorption-corrected rest-frame broad-band X-ray luminosities are
compared to the intrinsic column densities in Figure~\ref{lbx_nhe} for the
entire AGN sample (132 sources). Full symbols represent sources in sample A
(117 sources), and open symbols represent sources in sample B (15 sources).
Upper limits to the column density are adopted for sources with an observed
column density lower or consistent with the Galactic value. The upper limit
to the column density is derived from the Galactic value after applying the
correction for redshift (\nh$^{Gal}
\times (1+z)^{2.6}$). Median luminosity values, derived from the A
sample, for each class are shown with dashed lines. The absorption-corrected
rest-frame X-ray luminosities of AGN1 are on average more than ten times 
those of SF, while AGN2 show intermediate values. No correlation with the
intrinsic absorption measured in the X-rays is observed. The paucity of
heavily absorbed (\nh$\geq 10^{23}$\cm2) sources at luminosities below
10$^{44}$ \ergs, it is due to a selection effect because they would be too
faint in the X-rays to be detected above our selection threshold.

Note that the area (1\sqdeg) of this survey provides a large sample of
highly luminous (L$^{corr}_{0.5-10~keV}>10^{45}$~\ergs) AGNs (21 sources).
As commonly found in X-ray selected samples where unabsorbed type 1 AGNs
dominate the most luminous population~\citep{ueda03,akylas06}, we also find
a larger fraction of AGN1s at higher X-ray luminosities. The fraction of
AGN1s increases from 22\% at $L_{0.5-10~keV}^{corr}<10^{45}$ \ergs to 81\%
at $L_{0.5-10~keV}^{corr}\geq10^{45}$ \ergs. A detailed discussion on the
fraction of unobscured (AGN1s) and obscured (AGN2s+SFs) AGNs as a function
of X-ray luminosity and a comparison with previous results is reported in
Paper II. Paper II also discusses the uncertainty of these estimates due to
the large corrections in luminosity and column density for sources at
high-redshifts and the properties of absorbed QSOs (\nh$\geq$10$^{22}$\cm2
and $L_{2-10~keV}^{corr}>10^{44}$ \ergs). It is interesting to note the
lack of SF sources in sample A at high X-ray luminosities, $>$10$^{45}$
\ergs. This indicates that powerful AGNs always dominate in the IR and are,
therefore, identifiable as AGNs at those wavelengths. Thus, IR surveys can
potentially be used to identify and study the most luminous AGNs. There are
however two sources in sample B (XMDS seq: 403, 449) classified as SF at
high luminosities ($L_{2-10~keV}^{corr}>10^{45}$ \ergs).  Since they belong
to the B sample, they are not included in the analysis (open triangles in
Figure~\ref{lbx_nhe}). Deeper multi-wavelength observations are necessary to
better characterize these sources (classification and redshift) and verify
their high X-ray luminosities.

The mid-IR luminosity is compared with the column density measured in the
X-rays in Figure~\ref{lmir_nhe}. Similarly to the distribution of the X-ray
luminosities, the majority of AGN1s show predominantly high mid-IR
luminosities ($\geq$10$^{45}$\ergs), while SFs show mostly low mid-IR
luminosities ($<10^{44.5}$\ergs), and AGN2s show intermediate values
(10$^{44-45.5}$\ergs) overlapping with those of the other two classes. As
for the X-ray luminosities, the mid-IR luminosities of the objects in each
class do not show any obvious correlation with the intrinsic absorption
measured in the X-rays.

In the next sections, we will investigate whether the observed
multi-wavelength properties of the three classes can be simply explained by
different luminosities, dust extinction or orientation effects. These
parameters will be investigated separately for each AGN class by analyzing
the mid-IR and X-ray luminosity ratio, the dependency of the SEDs on the
X-ray luminosity, and of the IR colors on the luminosity and dust
extinction.

\section{The range of mid-IR over X-ray luminosity ratios}\label{lmir_lx_ratio}

The nuclear mid-IR continuum in AGNs is due to reprocessing of the AGN
emission by circumnuclear dust, e.g. the putative torus, and is thus 
function of both the AGN luminosity and the distribution of the obscuring
matter. The dust covering factor in AGNs can be investigated by comparing
the ratio between the mid-IR continuum (re-radiation) and the intrinsic
X-ray emission. A lower mid-IR/X-ray luminosity ratio is expected in highly
luminous AGNs if the dust inner radius recedes~\citep{lawrence91a} and the
torus height is constant, and in type 2 sources where the torus is seen
edge-on because of mid-IR absorption~\citep{granato97}. However, studies on
Seyfert 1, Seyfert 2 galaxies~\citep{lutz04,horst06}, and type 2
QSOs~\citep{sturm06} where the dust covering factor is estimated from the
$L_{mid-IR}/L^{corr}_{2-10~keV}$ luminosity ratio, where $L_{mid-IR}$ is
the monochromatic luminosity at 6 or 12~$\mu$m, have not found any dependency
on AGN type, absorption or X-ray luminosity. 

\begin{figure}
    \epsscale{1.0}
    \plotone{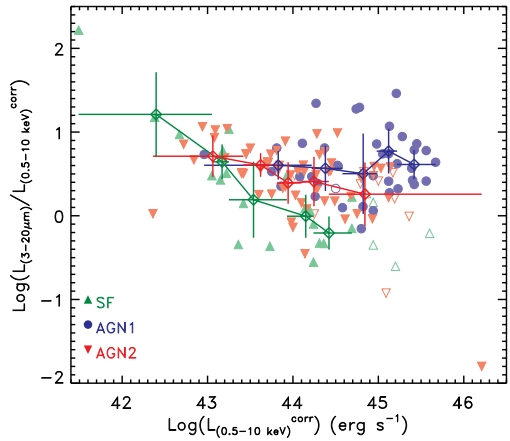}
    \caption{Ratio between the mid-IR and the absorption-corrected
        broad-band (0.5--10~keV) rest-frame X-ray luminosity as a function
        of the absorption-corrected broad-band (0.5--10~keV) rest-frame
        X-ray luminosity.  Symbols as in Figure~\ref{lbx_nhe}. The average
        ratios in bins of intrinsic X-ray luminosity derived for A sample 
        (full symbols) are shown as diamonds and connected
        by solid lines. The colors
        correspond to the three SED classes, AGN1: blue, AGN2: red, and SF:
        green.}
              \label{lmirlxc_lxc}
    \end{figure}

In Figure~\ref{lmirlxc_lxc}, we compare the mid-IR/X-ray luminosity ratio
with the absorption-corrected X-ray luminosity for the three AGN classes. In
order to investigate whether there is a trend with the X-ray luminosity or
the AGN class, we regrouped the data for each class in five bins with the
same number of sources per bin for each class, and derived the average
ratio. The average values are derived only for sample A (full symbols) and
are shown as diamonds connected by solid lines. For clarity, we show
the data of the single sources without uncertainties. Analogously to
previous studies~\citep{lutz04,horst06,sturm06}, we find that the
mid-IR/X-ray luminosity ratio is characterized by a wide dispersion and does
not depend on the AGN class or the X-ray luminosity. The only exception is
in an trend observed for the SF class to have a decreasing mid-IR/X-ray
luminosity ratio at increasing X-ray luminosities. Some models predict
smaller dust covering factors, or lower mid-IR over X-ray luminosity
ratios, in more luminous sources. However, according to these models, this
effect should be observed in all AGN classes and at extreme X-ray
luminosities, but the mid-IR/X-ray luminosity ratio in AGN1s and AGN2s is
constant and low ratios are also observed in SFs at moderate X-ray
luminosities ($\simeq 10^{43.5}$\ergs) as shown in Figure~\ref{lmirlxc_lxc}.

\begin{figure}
    \epsscale{1.0}
    \plotone{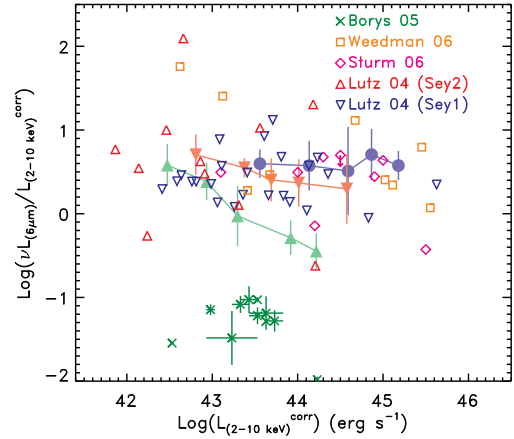}
    \caption{Ratio between the 6~$\mu$m and the absorption-corrected
       hard-band (2--10~keV) rest-frame X-ray luminosity as a function of
       the absorption-corrected hard-band (2--10~keV) rest-frame X-ray
       luminosity. Symbols refer to sources in the literature (reverse blue
       triangles  and red
       triangles: Seyfert 1s and Seyfert 2s from~\citet{lutz04}, magenta
       diamonds: type 2 QSO2s~\citep{sturm06}, orange squares: absorbed
       AGNs~\citep{weedman06}, and green crosses: sub-millimetre detected
       AGNs~\citep{borys05}).  The average ratios in bins of 
       X-ray luminosity derived for sample A are shown connected by solid
       lines as full blue circles, full reversed red triangles, and green
       triangles, for the three SED classes, AGN1, AGN2, and SF, respectively.}
              \label{lmirlxc_lxc_lit}
    \end{figure}

In order to further investigate whether there is a dependency with
luminosity, and whether previous studies did not cover a broad enough range
of parameters (X-ray luminosities, spectral types) to observe such a trend,
we produce a similar diagram using data from the literature. Instead of the
integrated mid-IR luminosity we use the monochromatic 6~$\mu$m luminosity
and the absorption-corrected hard (2--10 keV) X-ray luminosity because
available for all sources in the literature. The mid-IR over X-ray ratios as
a function of X-ray luminosities from various AGN
samples~\citep{borys05,sturm06,lutz04,weedman06} are shown in
Figure~\ref{lmirlxc_lxc_lit}. In addition to the literature data, we report
the average trend for the three AGN classes studied here. In this case, the
average values were derived using the monochromatic 6~$\mu$m luminosity and
the absorption-corrected hard (2--10~keV) X-ray luminosity as for the
literature sample. The resulting trends for our sample are consistent with the
results shown in Figure~\ref{lmirlxc_lxc}. The literature data show a large
dispersion, in agreement with our results, but no dependency on AGN type or
X-ray luminosity is observed. There is a group of sources
from~\citet{borys05} with unusual low ratios and moderate X-ray
luminosities. These sources are sub-millimetre detected AGNs whose
optical-IR emission is mainly dominated by the host-galaxy, similarly to the
SF class. The X-ray emission of these sources is faint, L${2-10
keV}^{corr}$$\simeq$10$^{42.5-44.2}$\ergs, and they are characterized by low
or moderate accretion rates, $dM/dt\simeq$0.01--0.80~\citep{alexander05a}.


The comparison with the ratios from the literature samples confirms that the
mid-IR/X-ray luminosity ratio in AGNs is characterized by a broad dispersion
and there is no dependency with the X-ray luminosity. At low and moderate
X-ray luminosities some AGNs are characterized by large mid-IR/X-ray ratios
because of contribution from the host galaxy to the mid-IR luminosity. In
SFs, the AGN mid-IR emission is generally fainter relative to the
absorption-corrected X-ray luminosity compared to AGN1s and AGN2s. The
host galaxy emission into the mid-IR can increase the mid-IR/X-ray ratios in
SFs and make them consistent or even larger than those observed in AGN1s and
AGN2s, but in some cases, the mid-IR/X-ray ratios can be 100 times lower, as
observed in~\citet{borys05} sample. However, it is still not clear whether
the observed low ratios are due to absorption of the mid-IR AGN emission or
are intrinsic. Lower intrinsic ratios might imply different dust properties
(geometry, distribution, optical depth). For example, a low mid-IR/X-ray
ratio could be indicative of lack of hot dust, as in the case the dust were 
located at a relatively large distance from the central heating source, e.g.
10 pc~\citep{ballantyne06}. In this case, a dependency of mid-IR/X-ray ratio
with the X-ray luminosity would not be expected.

\section{SEDs as a function of luminosity}\label{seds_vs_lx}

  \begin{figure*}
   \epsscale{2.0}
    \plotone{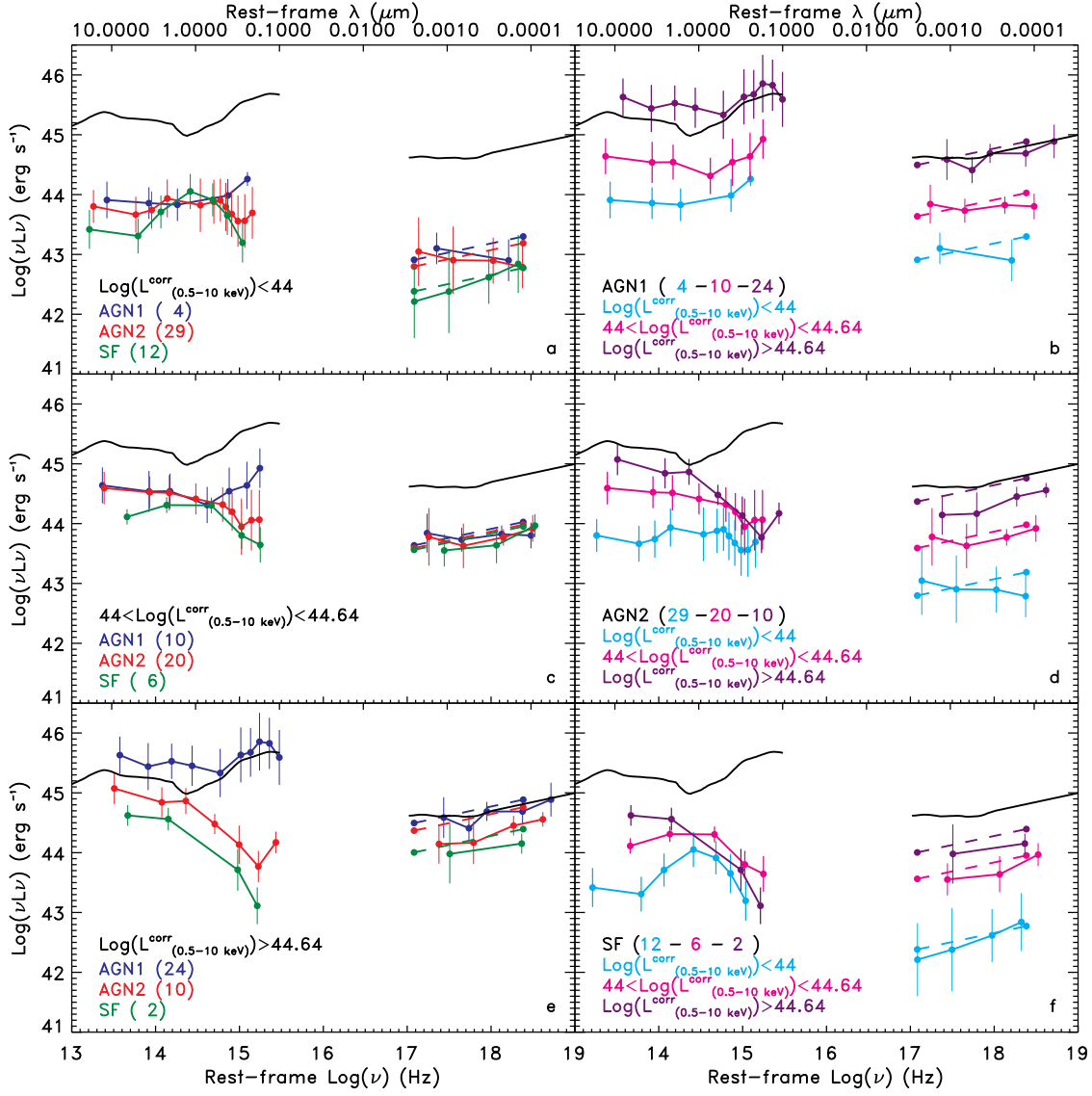}
   \caption{{\it Left panels:} Average rest-frame mid-IR-X-ray SEDs
     in Log($\nu$L$_{\nu}$) $vs$ Log($\nu$) of all AGNs with
     absorption-corrected broad-band X-ray luminosity lower than 10$^{44}$
     \ergs (panel a), 10$^{44-44.64}$~\ergs (panel c), and $>$10$^{44.64}$
     \ergs (panel e) are shown as filled circles connected by solid lines.
     Uncertainties correspond to the mean absolute deviation.
     Note that the X-ray SEDs are not corrected for absorption. The average
     SEDs for AGNs classified as SFs is shown in green, for AGN2s in red and
     for AGN1s in blue. The dashed lines represent the average
     absorption-corrected 0.5--10~keV rest-frame X-ray luminosities for each
     AGN class and luminosity group.  {\it Right panels:} Average rest-frame
     mid-IR-X-ray SEDs in Log($\nu$L$_{\nu}$) $vs$ Log($\nu$) of all AGN1s
     (panel b), AGN2s (panel d), and SFs (panel f) with absorption-corrected
     broad-band X-ray luminosity below 10$^{44}$
     \ergs\ (cyan), 10$^{44-44.64}$ \ergs\ (magenta), and $>$10$^{44.64}$
     \ergs\ (purple).  The number of sources used to derive the average SEDs
     is annotated. The black solid curve represents the median template for
     optically-selected QSOs~\citep{elvis94a}.}
              \label{rf_lum_seds_lx}
    \end{figure*}

In Section~\ref{agn_classes_seds}, we compared the normalized SEDs of the
three classes. Here, we investigate whether the average SEDs vary with the
AGN luminosity. We use the absorption-corrected broad band X-ray luminosity,
$L^{corr}_{0.5-10~keV}$, to separate each class (SF, AGN2, and AGN1) in
three groups, sources with 1) $L^{corr}_{0.5-10~keV}$ $<$ 10$^{44}$\ergs, 2)
$L^{corr}_{0.5-10~keV}$ = 10$^{44-44.64}$ \ergs, and 3) $L^{corr}_{0.5-10
keV}$ $>$ 10$^{44.64}$ \ergs. The luminosity values that define the three groups
correspond to the value that separates Seyferts and quasars,
$L^{corr}_{0.5-10~keV}=10^{44}$ \ergs, and $L^{corr}_{0.5-10~keV} =
10^{44.64}$ \ergs\ corresponds to the median luminosity of all sources with
$L^{corr}_{0.5-10~keV}$$>$10$^{44}$ \ergs. In group (1), 4 AGN1s, 29 AGN2s, and 12 SFs are
included; 10 AGN1s, 20 AGN2s, and 8 SFs in group (2), and 24 AGN1s, 10
AGN2s, and 2 SFs in group (3). The average SEDs of each class and group are
shown in Figure~\ref{rf_lum_seds_lx}. On the left side we show the SEDs
regrouped by X-ray luminosity and on the right side by class. Note that the
average absorption-corrected X-ray luminosities of each class within a group
can differ by up to a factor of 2. The Elvis's median QSO template is shown
for comparison. The average absorption-corrected X-ray luminosities are
shown as dashed lines at 0.5--10~keV in the rest-frame.

The optical-near-IR (0.4--2$\mu$m) luminosities of the three classes in the
lowest X-ray luminosity group (1) are all consistent within the 1$\sigma$
dispersion (panel a of Figure~\ref{rf_lum_seds_lx}). However, the SEDs are
characterized by different shapes, flat for the AGN1s consistent with AGN
emission at all wavelengths, and optically red and flat in the IR for the
AGN2s, consistent with being dominated by emission from the host-galaxy or
by a reddened AGN in the optical and from the AGN in the IR, and optically
red and blue in the IR for the SFs, consistent with being dominated by
emission from the host-galaxy at all wavelengths. In the X-rays, AGN1s
and AGN2s show soft spectra, consistent with no absorption, while SFs show a
hard spectrum. The sources in this class are more than 10 times less
luminous than Elvis's QSO median template at all wavelengths.

In group (2) (panel c), the optical-IR luminosities of the three classes
show similar properties to those in group (1). However, the difference in
luminosity at short ($\lambda<$0.4$\mu$m) and long ($\lambda>$2$\mu$m)
wavelengths in the three classes increases. The X-ray spectra are all
similar and consistent with the canonical slope ($\alpha_X$=0.7). The
difference in luminosity with Elvis's QSO median template is lower than in
group (1), but they are still from three to ten times less luminous.

In group (3) (panel e), the trend observed in the two previous groups is
confirmed, the optical-IR luminosities of the three classes separate
further, with the AGN1s being from ten to hundred times more luminous than
the SFs. In the X-rays, a large difference is observed between the
absorption-corrected luminosities and the observed values in AGN2s and SFs,
indicating large absorption, while no indication of X-ray absorption is seen
in AGN1s. The AGN1 luminosities are now consistent with Elvis's QSO median
template in the optical and X-ray, and higher in the IR.

The average SEDs in each class show also a change in the overall shape as
a function of X-ray luminosity (see right panels in
Figure~\ref{rf_lum_seds_lx}).

AGN1s (panel b) have overall similar infrared, optical and X-ray SEDs
at all luminosities, with only a slightly redder SED at $\lambda>1\mu$m and
less soft X-ray spectra at higher X-ray luminosities. The optical-IR
luminosities scale with the X-ray luminosities, both uncorrected and
corrected for absorption. 

The AGN2s (panel d) show a clear change in the optical-IR SED as the X-ray
luminosity increases, with the overall SED becoming redder. The most
plausible explanation for this change is an increasingly contribution from
hot dust continuum emission associated with the AGN, e.g the putative torus,
or less contribution from the host galaxy. In the lowest luminosity group,
the host galaxy still dominates, but at large luminosities its emission is
significantly lower than that of the AGN. It is difficult to characterize
the host galaxy emission in the mid-IR because it can change dramatically
for an elliptical or a late spiral or a starburst galaxy. In all cases
the host galaxy light peaks in the near-IR, $\sim$1-2~$\mu$m, where the AGN
light has a minimum~\citep{sanders89}. At longer wavelengths the host galaxy
emission can decrease or increase, according to the galaxy type, and the AGN
contribution increases. Thus, the total spectrum goes from blue if the host
dominates to flat if the two contributions are similar to red when the AGN
dominates. We can expect to see this trend for AGNs of increasing luminosity
if the host galaxy luminosity is fixed or does not scale with the AGN
luminosity. The X-ray spectra of the AGN2s change from being soft to hard as
the luminosity increases. It is interesting to note that the AGN mid-IR
luminosity does not scale with the absorption-corrected luminosity. Indeed
the mid-IR luminosity of the most X-ray luminous AGN2s is less than twice
that of those with intermediate X-ray luminosity, while the X-ray luminosity
is almost 10 times higher. This suggests mid-IR absorption in more
luminous sources or a lower dust covering factor.

The SF-like AGNs (panel f) show a similar change in SED shapes as the AGN2s.
In this class, the near-IR ($\sim$1~$\mu$m) luminosity remains approximately
the same in the three groups, however the absorption-corrected X-ray
luminosity is about 40 times larger in the most X-ray luminous SFs than in
the less X-ray luminous ones. The decrease in mid-IR luminosity relative to
the absorption-corrected X-ray luminosity with increasing X-ray luminosities
discussed in Section~\ref{lmir_lx_ratio} is also visible here. This result
is even more remarkable considering that the host galaxy contributes to the
mid-IR luminosity.

The decreasing mid-IR luminosity compared to the absorption corrected X-ray
luminosity observed in the most luminous AGN2s and in the SFs is equivalent
to the low mid-IR over X-ray ratios found at high X-ray energies in SFs
discussed in Section~\ref{lmir_lx_ratio} and shown in
Figures~\ref{lmirlxc_lxc} and~\ref{lmirlxc_lxc_lit}. From the analysis in
the previous section, we concluded that the deficiency of mid-IR light is
not dependent on the AGN X-ray luminosity. However, when the AGN X-ray
luminosity is low the host galaxy might be luminous enough to compensate the
mid-IR deficiency, preventing low mid-IR over X-ray ratios to be observable
in less luminous AGNs. Assuming that a molecular torus surrounds the nuclear
source, the deficiency of mid-IR light relative to the X-ray luminosity
observed in SFs and some AGN2s can be explained by lack of hot dust, larger
optical depths, or large angles of the line of sight from the torus
axis~\citep{granato94,granato97}.

\section{The origin of AGN IR colors}\label{agn_colors}

   \begin{figure*}
    \epsscale{2.2}
    \plottwo{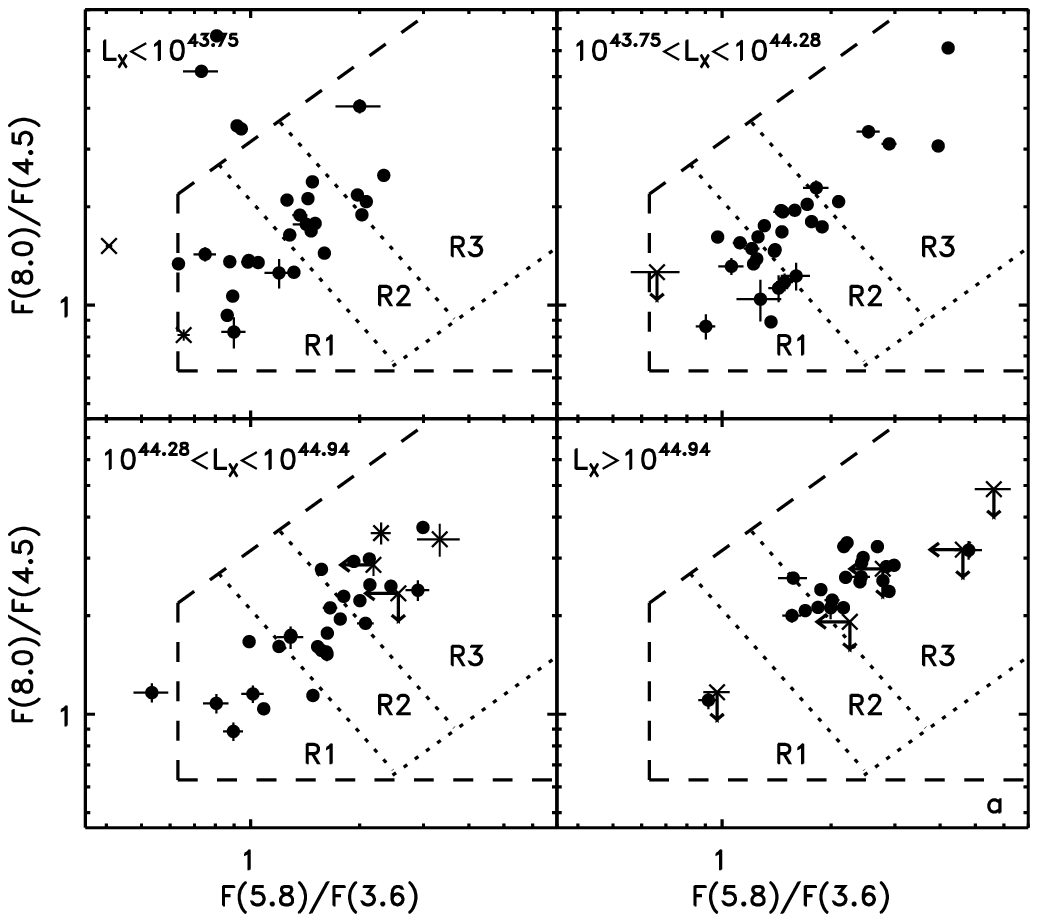}{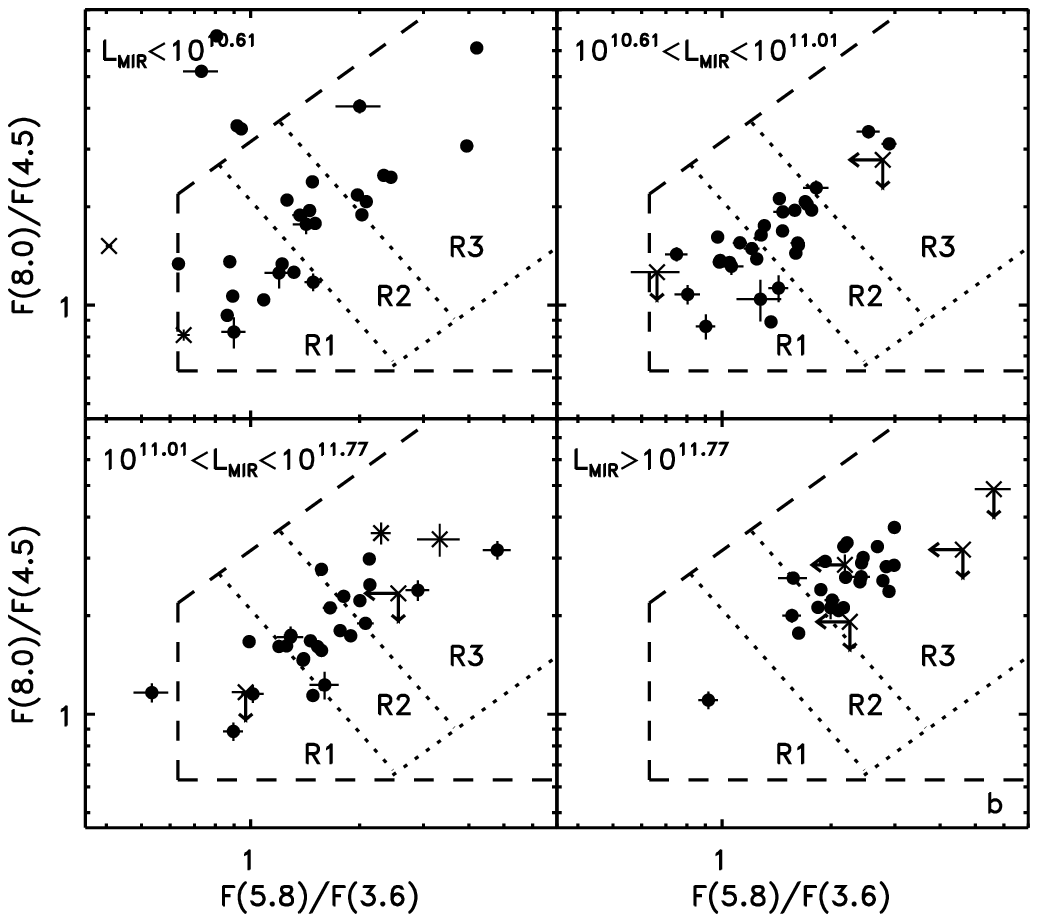}\\
    \epsscale{2.2}
    \plottwo{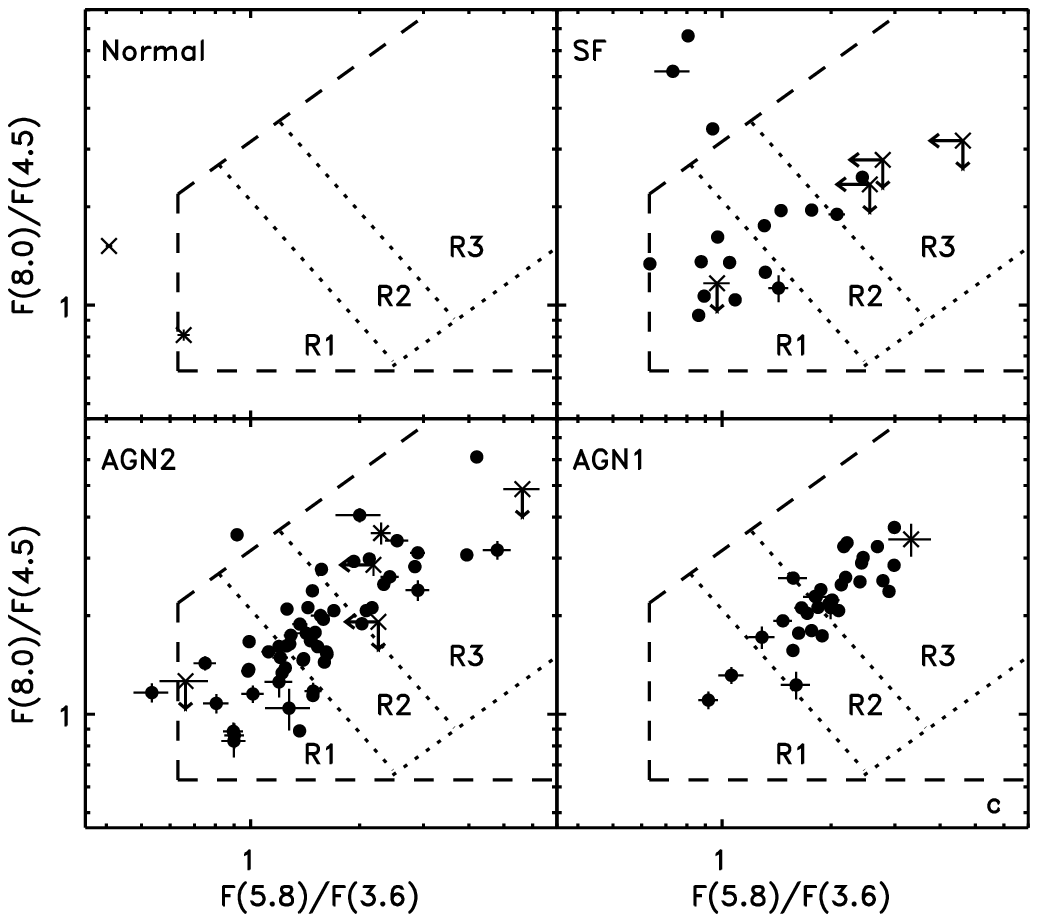}{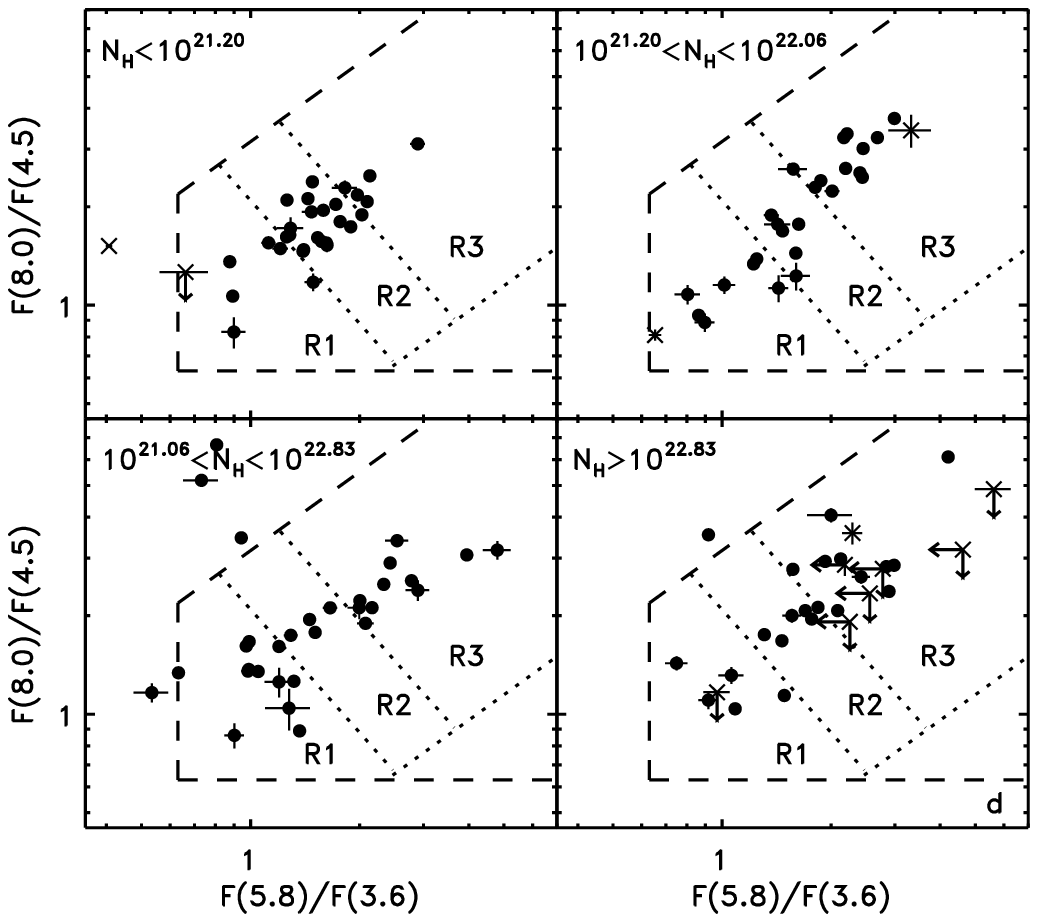}\\
   \caption{IRAC colors, F(8.0~$\mu$m)/F(4.5~$\mu$m) $versus$
      F(5.8~$\mu$m)/F(3.6~$\mu$m), as in~\citet{lacy04} for our sample. Full
      symbols represent sources in sample A and open symbols
      represent sources in sample B. Each figures shows
      sources with different
      absorption-corrected X-ray luminosity (a), mid-IR (3--20
      ~$\mu$m) luminosity (b), SED class (c), and column density as
      estimated in the X-rays (d). Each figure is divided
      in 4 panels with sources in 4 different sub-groups as annotated. Left
      pointing arrows represent 5$\sigma$ upper limits at 5.8~$\mu$m,
      downward pointing arrows represent 5$\sigma$ upper limits at
      8.0~$\mu$m. The region enclosed by the dashed line corresponds to the
      region identified by~\citet{lacy04} where the majority of
      IRAC-detected AGNs lies. The dotted lines define three sub-regions
      within the Lacy's region, labeled R1, R2, and R3, where the same
      number of sources in our sample lies. The number of sources in each
      group and the fraction in each region are listed in
      Table~\ref{lacy_fraction}.}
              \label{lacy_plots}
   \end{figure*}

\citet{lacy04} identified a region in IRAC color-space (3.6--8.0~$\mu$m) where
most of the AGNs detected in all four IRAC bands are distributed~\citep[see
also ][]{barmby06,stern04,hatziminaoglou05a}. AGNs occupy preferentially
this region because of their red and almost featureless IR SED. Only five
out of 87 sources in sample A with detections in all four IRAC
bands are not within the `AGN wedge' as defined by~\citet{lacy04}. For
fifteen additional objects, only an upper limit in one or more IRAC bands is
available. We can therefore conclude that Lacy's criterion is satisfied by
80\% of our X-ray sources in sample A observed with IRAC (this fraction is
obtained with the pessimistic assumption that all the sources with IRAC
limits would fall outside Lacy's wedge) or by a fraction close to 95\%
(97/102, if we assume that all the sources with upper limits lie within the
wedge). Thus, this criterion appears to be fairly complete for a relatively
bright X-ray sample as the XMDS sample. Using the properties (SED class,
luminosity, and absorption) derived for the AGNs in the XMDS, we
investigate how these affect the IRAC colors and make them so distinctive
for AGNs.

In the hypothesis that different AGN types and column densities correspond
to tori seen at different angles with respect to the line of sight and that
the mid-IR radiation is dominated by the torus emission, a dependence of the
IR colors on the AGN class (type 1 $vs$ type 2) and on the absorption in the
X-rays is expected. In Figure~\ref{lacy_plots}, we reproduce Lacy's color
diagram plotting separately sources in different intervals of
absorption-corrected X-ray luminosity (panel a), mid-IR (3--20~$\mu$m)
luminosity (panel b), SED classification (panel c), and column density
(panel d). IRAC colors are constrained for 107 out of the 117 sources in
sample A, and 102 satisfy Lacy's color criteria: all AGN1s, and most AGN2s
(96\%) and SFs (82\%). We divided the area defined by~\citet{lacy04} in
three sub-regions, R1, R2, and R3, that contain sources with increasing flux
ratios or redness and, approximately, the same number of sources ($\sim$34).
The equation of the line that divides regions R1 and R2 is
$Log(F(8.0)/F(4.5))=0.32-1.25\times Log(F(5.8)/F(3.6))$, and the equation of
the line that divides regions R2 and R3 is
$Log(F(8.0)/F(4.5))=0.661-1.25\times Log(F(5.8)/F(3.6))$. In order to
investigate whether any AGN property (class, X-ray luminosity, mid-IR
luminosity, and X-ray absorption) is correlated with the mid-IR
redness, we derived the fraction of sources with a specific property in each
region. The number of sources per region, including upper limits or
detections only, are reported in Table~\ref{lacy_fraction}.

The position of the sources in the diagram is strongly correlated with
the SED class. The fraction of AGN1s increases from R1 to R3, while most of
the SFs are in R1 and most of the AGN2s in R2. The observed dependency is
the opposite of what expected by AGN tori models where redder SEDs are
predicted in more obscured sources. A likely explanation for this
correlation is the increasing host-galaxy contribution from AGN1s to SFs. A
correlation with the IRAC colors is also observed as a function of X-ray
and mid-IR luminosities, with the fraction of the most luminous
sources increasing and that of the less luminous decreasing going from R1
and R3. This correlation is consistent with less host contribution in the IR
in more luminous AGNs. The least correlated parameter with the IRAC colors
is the column density measured in the X-rays. However, all sources in the A
sample outside~\citet{lacy04} region are X-ray absorbed.

\section{Radio luminosity $vs$ X-ray and mid-IR luminosities}\label{radio}

In star-forming galaxies, the radio and far-infrared (FIR;
42.5--122.5~$\mu$m) emission are strongly
correlated~\citep{dickey84,dejong85}. The majority of AGNs (radio-quiet
AGNs), about 90\%, depending on the sample selection, follow the same
correlation as star-forming galaxies, and a minority, i.e. radio galaxies
and radio-loud quasars, shows an excess of radio emission compared to
radio-quiet objects with similar FIR emission~\citep{sopp91,roy98a}. The FIR
over radio flux ratio is thus a good diagnostic tool to identify AGN
activity. In recent studies, this correlation has been used by replacing the
FIR emission with the 24~$\mu$m emission because more easily available for
faint high-$z$ sources thanks to
\spitzer~\citep{appleton04,higdon05,donley06}, although the correlation
between the 24~$\mu$m and the FIR emission is characterized by a large
dispersion~\citep{brandl06,barger06}. The radio continuum is also correlated
with the X-ray emission in AGNs~\citep[e.g.; ][]{brinkmann00,simpson06}. A
correlation between the radio and the X-ray emission is also observed in
star-forming galaxies, however the X-ray luminosity is on average 400 times
lower than in AGNs of similar radio luminosity~\citep{ranalli03}. Here, we
investigate whether our data are consistent with the well known FIR-radio
and X-ray-radio correlations and whether radio data can help identifying
heavily obscured AGNs.

   \begin{figure}
    \epsscale{1.0}
    \plotone{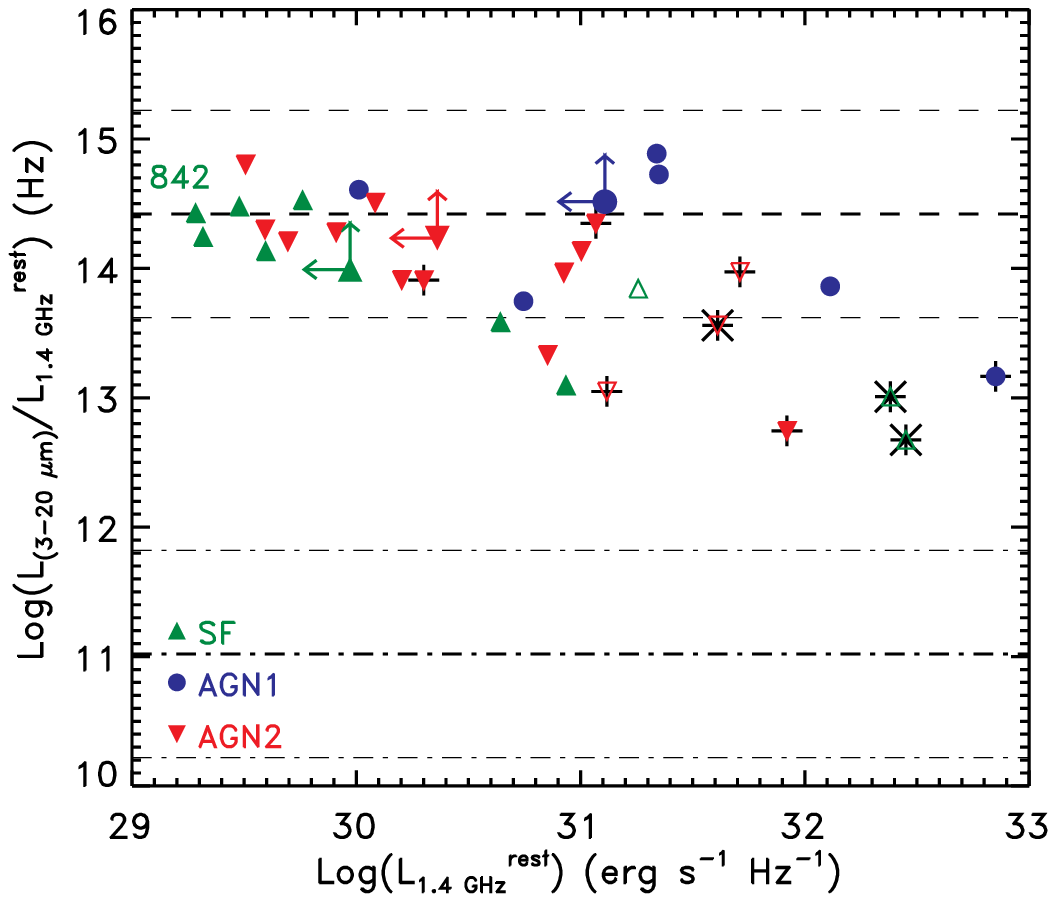}
   \caption{Ratio between the rest-frame mid-IR (3--20 ~$\mu$m) luminosity
            and the rest-frame 1.4~GHz radio luminosity $vs$ the rest-frame
            1.4~GHz radio luminosity. Symbols as in Figure~\ref{lbx_nhe}. 
            The thick dashed line represents the L(mid-IR)-L(Radio)
            correlation derived from the L(FIR)-L(Radio) correlation
            observed for radio-quiet AGNs and star-forming
            galaxies~\citep{sopp91,ranalli03}. The thin dashed lines
            represent the dispersion associated with the conversion from the
            FIR luminosity to the mid-IR luminosity. The thick dotted-dashed
            line represents the L(mid-IR)-L(Radio) correlation derived from
            the L(FIR)-L(Radio) correlation observed for radio-loud
            AGNs~\citep{sopp91}. The thin dashed lines represent the
            dispersion associated with the conversion from the FIR
            luminosity to the mid-IR luminosity. Plus signs represent 9
            sources classified as radio-loud based on the radio over optical
            or radio over infrared flux ratio, crosses are over plotted on
            the 3 sources that satisfy both criteria (see text). 
            The upward-leftward arrows represent
            the median radio luminosity and luminosity ratio obtained from
            the non-detected sources in each group.}
               \label{lrad_lmir}
    \end{figure}
   \begin{figure}
    \epsscale{1.0}
    \plotone{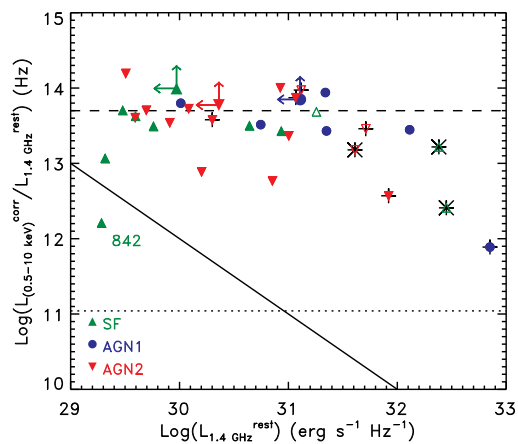}
   \caption{Ratio between the absorption-corrected rest-frame broad (0.5--10
            keV) X-ray luminosity and the rest-frame 1.4~GHz radio
            luminosity $vs$ the rest-frame 1.4~GHz radio luminosity. Symbols
            as in Figure~\ref{lrad_lmir}. The dashed line represents the
            L(X-ray)-L(Radio) correlation derived for radio quiet AGNs in
            the ROSAT-FIRST survey~\citep{brinkmann00}. The dotted line
            represents the L(X-ray)-L(Radio) correlation derived for
            star-forming galaxies~\citep{ranalli03}. The
            black solid
            diagonal line corresponds to a fixed X-ray luminosity of
            10$^{42}$ \ergs. }
              \label{lrad_lx}
    \end{figure}

There are 32 sources in our sample detected at 1.4~GHz, six belong to the B
sample (three with dubious optical counterparts and three with poorly
sampled SEDs). Upper limits to the radio flux are available for 68 sources
(2 are in the B sample). Thus, there are 26 radio-detected and 66
non-detected sources in sample A. Radio data can be used to identify the
presence of an AGN by looking for an excess of radio emission compared to
the optical and/or the IR flux. According to~\citet{stocke92}, radio-loud
AGNs are characterized by a radio over optical ratio, $R=F_{5
GHz}/F_{2500\AA}$, greater than 100\footnote{The definition of radio-loud
based on the parameter $R$=F$_{5 GHz}$/F$_{2500\AA}$ depends on whether the
parameter $R$ is k-corrected and on what spectral indexes are assumed in the
optical and in the radio. Typical values are Log$R<$0.5 for radio-quiet and
Log$R>$2 for radio-loud objects~\citep{stocke92}.}. We derive the radio flux
at 5 GHz assuming a spectral index $\alpha_R$=0.8 (F$_{\nu}\propto
\nu^{-\alpha_R}$), and the optical flux by interpolating the observed data
at the wavelength corresponding to 2500\AA\ rest-frame. Seven sources have
$R\geq$100 (XMDS seq: 18, 58, 111, 120, 403, 747, and 1219) of which two
(XMDS seq: 403, and 1219) are optically blank and five are in sample B (XMDS
ID: 111, 120, 403, 747, and 1219). We also calculated the parameter
$q$ defined as $Log(F_{24\mu m}$/F$_{20cm})$, and identified a radio excess
in all sources with $q<$0.0~\citep{donley05}. Five sources satisfy this
criterion, XMDS seq: 403, 738, 747, 1219, and 1246. Three sources are selected
by both criteria (XMDS seq 403, 747, and 1219), for two (XMDS seq 111, and 120)
the $q$ parameter can not be derived because there is no IR coverage. In
summary, the presence of an excess of radio emission remains uncertain in
six sources for which only one of the two criteria is satisfied or can be
applied, XMDS seq: 18, 58, 111, 120, 738, and 1246, and it is confirmed by both
criteria in 3 sources, XMDS seq: 403, 747, and 1219. 

In Figures~\ref{lrad_lmir} and~\ref{lrad_lx}, we investigate the correlation
between the radio luminosity and the mid-IR and X-ray luminosity by
comparing the ratio between the rest-frame integrated mid-IR (3--20~$\mu$m)
and absorption-corrected rest-frame broad-band (0.5--10~keV) X-ray
luminosities and the rest-frame radio luminosity at 1.4~GHz with the
rest-frame radio luminosity at 1.4~GHz for the 32 X-ray sources that are
detected at radio wavelengths. We show the luminosity ratio to remove the
redshift effect. The 9 radio-loud objects are shown as black plus signs, and
crosses are over-plotted on the 3 more secure radio-loud AGNs. The 6 sources
in sample B are shown as open symbols. Although these sources are
classified as radio-loud by the criteria described above, based on their
radio-to-MIR and radio-to-X-ray luminosity ratios, it would be more
appropriate to consider them as radio-intermediate AGNs. For clarity we do
not report sources that are not radio-detected, but for each class we show
the median upper limit to the radio luminosity and to the luminosity ratio
obtained from the non-detected sources.

In Figure~\ref{lrad_lmir}, we test whether our sample shows the same
correlation between the far-IR (60~$\mu$m) luminosity and the radio
luminosity at 5 GHz as observed in local radio-quiet and radio-loud
AGNs~\citep{sopp91}. We converted the monochromatic luminosity at 5 GHz to
that at 1.4~GHz assuming a radio spectral index $\nu_R$=0.5 ($I\propto
\nu^{-\alpha_R}$) as in~\citet{sopp91} and the monochromatic luminosity at
60~$\mu$m to the mid-IR luminosity using the following equation, $L_{3-20\mu
m} = (0.4\pm0.8)\times \nu L_{60\mu m}$, derived from the templates that fit
the SEDs of our sample. The expected relationships are: $Log(L_{1.4~GHz}) =
Log(L_{3-20\mu m})-11.02\pm0.8$ for the radio-loud sources and $Log(L_{1.4
GHz}) = Log(L_{3-20\mu m})-14.42\pm0.8$ for the radio-quiet ones, where
$L_{1.4~GHz}$ is in erg s$^{-1}$ Hz$^{-1}$ and $L_{3-20\mu m}$ in \ergs.
The two relationships are shown in Figure~\ref{lrad_lmir} as thick dashed
(radio-quiet) and dot-dashed (radio-loud) lines. The thin lines represent
the dispersion associated with the conversion from $\nu L_{60\mu m}$ to
$L_{3-20\mu m}$. Most of the radio-quiet sources in our sample follow the
expected correlation. The Spearman correlation rank between the radio and
the mid-IR luminosities is $\rho_S$=0.84 and the probability is 1$\times10^{-07}$.
The lowest luminosity ratios are observed for the sources considered
radio-loud based on the optical/radio or the IR/radio flux ratios, however
they all show higher ratios than expected for radio-loud sources
(dot-dashed line in Figure~\ref{lrad_lmir}).

We find a similarly strong correlation between the rest-frame radio
luminosity at 1.4 GHz and the absorption-corrected rest-frame broad-band
(0.5--10~keV) X-ray luminosity (Spearman rank correlation coefficient
$\rho_S$=0.91 and probability of its deviation from zero
1.9$\times$10$^{-10}$). We compared the measured relationship with that
found for radio-quiet AGNs~\citep{simpson06,brinkmann00} (dashed line).  The
expected relationship has been derived from the correlation observed between
the monochromatic luminosities at 5 GHz and at 2~keV of a sample of
radio-quiet ($Log R\leq 1$) sources from the ROSAT-FIRST
sample~\citep{brinkmann00}. The radio luminosity has been converted to
$L_{1.4~GHz}$ assuming a radio spectral index of 0.7 and the X-ray
luminosity at 2~keV has been converted to
$L_{0.5-10~keV}$ assuming a spectral index $\alpha_X$=0.8 as assumed
in~\citet{brinkmann00}. The final relationship is $Log(L_{1.4~GHz}) =
Log(L_{0.5-10~keV})-13.91$~\citep[see also][]{simpson06} and is shown as a
dashed line in Figure~\ref{lrad_lx}. The source in our sample show
luminosity ratios consistent with those expected based on this
relationship.

The agreement with correlations derived from samples selected in different
energy bands, and at different flux limits, reinforces the validity of these
correlations. Since the radio luminosity can be considered an isotropic
indicator of the nuclear unabsorbed X-ray luminosity for radio-quiet AGNs,
these correlations can be used to probe the unabsorbed X-ray
luminosity of radio-quiet AGNs and to identify absorption effects in X-ray
observed AGNs, similarly to the analysis based on the \oiii\
luminosity~\citep[see
e.g.][]{mulchaey94,alonso-herrero97,turner97,maiolino98,risaliti99,bassani99}.

The XMDS sample offers an opportunity for such an application. Source XMDS
ID 842 does not follow any of the correlations between the X-ray and the
radio luminosities expected for AGNs or star-forming galaxies. Because of
its low X-ray luminosity, lower than 10$^{42}$ \ergs, the evidence for AGN
activity from the X-rays in this sources it is only based on the hardness of
its X-ray spectrum. However, its radio over X-ray luminosity ratio is
$\sim$ 10 times larger than what expected for normal star-forming
galaxies~\citep{ranalli03}, supporting the assumption that this source is an
AGN. This behaviour suggests that the X-ray luminosity might be
underestimated by a factor of 40, a hypothesis that would indicate that the
source is Compton-thick and the observed X-ray emission is only scattered
light. This source will be observed with the $\gamma$-ray observatory,
INTEGRAL, as part of a 2 Msec survey of the XMM-LSS field~\citep{virani05}.
We estimate the luminosity and expected flux in the energy range 20-100 keV,
where INTEGRAL is sensitive, to determine whether INTEGRAL will be able to
detect it. Assuming that the absorption-corrected rest-frame 0.5--10~keV
luminosity is 40 times higher, the derived luminosity is $L_{20-100
keV}$=2.9$\times$10$^{43}$ \ergs, which corresponds to an absorption
corrected flux $F_{20-100 keV}^{corr}\simeq7\times10^{-12}$ \ergcm2s, and an
observed flux $F_{20-100 keV}\simeq6.8\times10^{-12}$ \ergcm2s assuming a
column density of 10$^{24}$~\cm2. So far, INTEGRAL has not detected AGNs at
these flux levels~\citep{beckmann06,bassani06}, however the long XMM-LSS
exposure might be able to reveal the heavily obscured AGNs in source XMDS seq
842 (aka Arp 54).

\section{Discussion}\label{discussion}

\subsection{AGN identification through SED fitting}\label{agn_identification}

In this work, we have characterized the SEDs of a hard X-ray selected sample
of AGNs. As previously found~\citep[see
e.g.][]{alexander01a,alexander01b,alexander02,fadda02,perola04,szokoly04,franceschini05,polletta06},
the multi-wavelength SEDs of X-ray selected AGNs show a wide
variety of properties. We divide the SEDs in three broad classes, AGN1,
AGN2, and SF. The fraction of AGNs that fall into the various SED classes
likely depends on the sample selection, i.e. spectral window and flux limit.
For example, an optically selected sample will have a larger fraction of
AGN1s and an IR-selected sample will be biased against AGNs in the SF class.
A different X-ray flux limit or a selection in a different X-ray band can also
alter the fraction of sources in the various classes. For example, 
\citet{franceschini05} finds 39\% AGN1s, 23\% AGN2s and 38\% SFs in a 
sample of AGNs with broad X-ray fluxes $F_{0.5-8 keV}>10^{-15}$\ergcm2s. If
the classification of the sources fitted with a Mrk 231 template
in~\citet{franceschini05} is modified from AGN1 to AGN2 as assumed here, the
fractions become 15\% of AGN1s, 42\% of AGN2s, and 36\% of SFs. The
AGN1 fraction is lower than ours (15 $vs$ 33\%), the fraction of AGN2s is
slightly lower than ours (42 $vs$ 50\%) and the fraction of SFs is twice
ours (36 $vs$ 17\%). These differences are likely due to our higher
X-ray flux limit and requirement of detection in the hard X-ray band
(2-10 keV). There is indeed an indication for the fraction of AGN1s to
decrease and the fraction of SFs to increase at fainter X-ray fluxes (see
Paper II for a more detailed analysis). The trend of having less AGN1s and
more SFs at lower X-ray flux levels has to be confirmed and better
quantified using similarly selected samples with analogous classification
covering a broad range of fluxes.

Although hard X-ray samples, as the one used here, provide the least biased
AGN samples, some AGN, e.g. Compton-thick AGNs are still missed. However, we
do not expect the SEDs of Compton-thick AGNs to fall in a separate class.
Known Compton-thick AGNs, both in the local universe and at high redshifts,
show SEDs that are consistent with the AGN2 (e.g. Mrk 231, NGC 1068) and SF
classes (e.g. NGC 4945, NGC 6240). From the SED analysis of a sample of
X-ray selected Compton-thick AGNs,~\citet{polletta06} estimates that 60\% of
Compton-thick AGNs fall in the SF class, and 40\% in the AGN2 class. Thus,
through SED fitting we can identify a large fraction of AGNs, even
Compton-thick AGNs. The SED fitting technique provides an efficient AGN
identification tool. The main advantage is that it is based on data that can
be more easily obtained than X-ray data and spectroscopic data. Moreover,
this technique identifies the same or an even higher fraction of AGNs than
the spectroscopic classification. Based on the sample of 49
spectroscopically identified XMDS sources, we estimate that the SED fitting
technique is able to identify AGN activity in 8 out of 11 (73\%) AGNs that
are spectroscopically classified as star-forming galaxies, in 7 out of 12
(58\%) AGNs that are spectroscopically classified as type 2 AGNs, and in 25
out of 26 (96\%) AGNs that are spectroscopically classified as type 1 AGNs.
Overall, we identify AGN activity in 40 out of 49 (82\%) hard X-ray selected
AGNs with available spectroscopic data and in 83\% (50\% of AGN2s and 33\%
of AGN1s) in the entire X-ray sample, while optical spectroscopy reveals an
AGN in 38 sources (78\%). The SED fitting technique can be applied to
sources that are too faint at optical wavelengths for optical spectroscopic
observations, or with emission lines that are diluted by the host galaxy
light or affected by obscuration~\citep{moran02,barger01,rigby06}, and to
sources in the redshift desert. There is evidence that these source might
represent a significant fraction of AGN samples. Thus, SED fitting
represents a powerful and efficient tool to identify AGN activity in samples
for which optical spectroscopic observations are unavailable or impossible
to perform.

However, this technique has also some limitations. Almost half of type 2
AGNs and a third of star-forming galaxy-like AGNs might still be missed and
not identified as AGNs. Moreover, the SED classification can be misled by
the more dominant host-galaxy light at optical-near-infrared wavelengths as
in many Seyfert 1 galaxies and thus classify a type 1 AGN as AGN2 or SF. 
These limitations might be overcome if well sampled multi-wavelength SEDs of
large AGN samples in various classes with spectroscopic data become
available. With such data the AGN templates can be improved and expanded
and the SED fitting technique can be refined.

In the next section we discuss the nature of the AGN that are not identified
by the SED technique and by optical spectroscopy.

\subsection{The nature of elusive AGNs}\label{sf_nature}

There are several examples of AGNs in the literature whose optical spectra
or broad-band SEDs resemble those of normal galaxies and are, therefore, not
identified as AGNs unless probed by X-ray or radio data. We refer to these
sources as elusive AGNs~\citep{maiolino03}. These sources belong to the SF
class and might have many properties in common with known categories of
elusive AGNs found in the literature, e.g the optically dull discovered in
optical samples~\citep{maiolino03,elvis81,rigby05}, the X-ray Bright
Optically Normal Galaxies discovered in X-ray samples~\citep[XBONGs;
][]{comastri02}, the type 3 QSOs found in IR-selected
samples~\citep{martinez06,leipski07}. The rich taxonomy of these categories
reflects mostly the wavelengths at which they were discovered, while the
scenarios proposed to explain their nature are similar.

The properties of these elusive AGNs are usually explained by two scenarios:
1) heavy obscuration by dust with a large covering
factor~\citep{marconi00,fabian04,dudley97}, and/or 2) a radiatively
inefficient accretion flow (RIAF)~\citep{yuan04}. Large dust covering
factors obscure the optical emission from the inner regions, thus no narrow
emission lines are observed. The inefficient accretion explains the lack of
optical and ultraviolet radiation from the accretion disk and of emission
lines from the broad and narrow line regions. The existence of unabsorbed
XBONGs~\citep{severgnini03,page03} in the X-rays disfavors the former
hypothesis, although it is likely that both scenarios might be valid. Based
on IR and sub-millimetre observations of some of these elusive
AGNs~\citep{borys05,alexander05a,alexander05b,martinez06}, it has been
proposed that the obscuring dust is associated with a starburst and that
these AGNs might be in an early phase of their growth.  These objects might
have already formed the bulk of their stellar mass, but the central BH has
not fully grown and they are accreting at a low
rate~\citep{alexander05b,borys05}. This last scenario combines the two
described above. In summary, all these classes of AGNs are either explained
by heavy absorption associated with dust with a large covering factor and/or
by an inefficiently accreting black hole.

We examine in more detail the properties of 3 SFs for which spectroscopic
data are also available and fail in identifying the AGN, XMDS seq 439, 487,
and 842. Our goal is to find out what they have in common that might explain
their elusiveness and whether we can identify their AGN activity by other
means in absence of X-ray data. All three sources are fitted with spiral
templates, however XMDS seq 487 shows a small excess of emission in the
near-IR with respect to the template. Such an excess is likely due to
AGN-heated hot dust. Indeed, a fit with an AGN1 template is one of its
secondary solutions. All three sources are absorbed in the X-rays
(\nh$\gtrsim$2$\times$10$^{22}$
\cm2) and their X-ray luminosities cover a broad range, from
10$^{41.5}$\ergs\ to 10$^{44.4}$\ergs. The MIR-to-radio luminosity ratios of
two sources are consistent with those observed in radio-quiet AGNs and
normal galaxies, however XMDS seq 439 shows a lower than expected ratio
($\simeq$13.1, compared to the expected 14.4; see Figure~\ref{lrad_lmir}).
This low ratio indicates radio emission in excess, and thus strongly
suggests AGN activity. In summary, by using the MIR-to-radio luminosity
ratio and the secondary solutions of the SED fits, we can identify 
AGN activity in 2 of these 3 sources. The AGN nature of the third source,
XMDS seq 842, is actually not confirmed (see Sections~\ref{sfg}
and~\ref{radio}).

Overall, we do not find any evidence for inefficient accretion in SFs.
Indeed they can be intrinsically powerful X-ray sources and their
X-ray/radio luminosity ratios satisfy the same correlations observed for
AGN1s and AGN2s. The results from this study on these sources favor
obscuration as the cause of the observed low mid-IR/X-ray luminosity ratios
and elusive SEDs. However, an analysis of a larger sample with a broad range
of luminosities and a rich multi-wavelength data set and, especially, more
discriminating model predictions for various dust distributions and
orientations, linked with the X-ray properties, are necessary to investigate
whether our results are simply consistent with an orientation effect, or
indicate higher dust covering factors or larger dust inner radii.

\subsection{The missing AGN population}\label{missing_agn}

The cosmic X-ray background is still unresolved at high energies ($\geq$10
keV). The most recent background synthesis models predict that the missing
population is made of sources at intermediate redshifts ($z\sim$1), with
hard X-ray spectra ($\Gamma\sim$1.4), and moderate luminosities
($\sim$10$^{43}$\ergs)~\citep{worsley05,gilli06}. \citet{simpson06} suggest
that the missing AGN population would be detected at radio wavelengths with
fluxes of a few tens of ~$\mu$Jy. The expected radio luminosity of an AGN
with an X-ray luminosity of 10$^{43}$\ergs\ derived using the correlation
shown in Figure~\ref{lrad_lx} is about 10$^{29}$\ergs\ Hz$^{-1}$. The
observed radio flux, assuming a redshift of 1 as predicted for the bulk of
this missing AGN population, would be $\simeq$ 2.2 ~$\mu$Jy and the observed
0.5--10~keV X-ray flux would be $\simeq$ 2.35$\times$10$^{-15}$\ergcm2s if unobscured
and reduced by 20\%, 50\%, or 80\% if obscured by a column density of
10$^{22}$, 10$^{23}$, or 10$^{24}$\cm2, respectively. The predicted radio
fluxes are lower than the limits reached by the deepest radio surveys.
Indeed, the r.m.s. of the deepest VLA surveys is 2.7~$\mu$Jy in the
0.4\sqdeg\ SWIRE/\chandra\ field~\citep[][; Owen et al., in
prep.]{polletta06}, and 5.3~$\mu$Jy in the 0.04\sqdeg\ GOODS-North
field~\citep{morrison06}. Thus, the faint radio population is not a good
candidate for the missing AGN population. Similar conclusions have been
recently derived from stacking the X-ray data of faint radio-selected
sources in HDFN~\citep{barger06}.

\section{Summary}\label{summary}

Using the large data set from the XMM-LSS, SWIRE, CFHTLS, and VVDS surveys
in the XMDS field (Paper I), the properties (SEDs, luminosity and
absorption) of a hard X-ray selected sample of 136 X-ray sources are
investigated. Photometric redshifts are estimated and spectral types are
determined using the Hyper-z code and a library of galaxy and AGN templates
using optical and infrared data (Section~\ref{photoz}). In this work, we
investigate the role of the main parameters responsible for the observed
properties of AGN SEDs, host galaxy contribution, gas obscuration, dust
absorption, and AGN power. Our analysis is based on multi-wavelength data,
on the characterization of the average SEDs, and on the comparison of the
luminosities emitted in the main energy windows, optical, mid-IR, X-rays and
radio. Out of 136 hard X-ray sources, 132 are AGNs, two are clusters and two
are star-forming galaxies. From the AGN sample, we select a sub-sample of
117 AGNs with high quality SEDs. The main results obtained from the analysis
of this sample are summarized below.

\begin{itemize}
\item {\bf AGN SED classification} We fit the optical-IR (from 0.37 to
24$\mu$m) SEDs of the whole sample (134 sources, excluding the two clusters)
to estimate photometric redshifts and classify the sources. Spectroscopic
data are available for 33 sources. We obtain photometric redshifts with
$\sigma$(1+$z$)=0.11 and 6\% of outliers. The AGN sample is divided in three
classes based on their SEDs, type 1 AGNs (AGN1s) (33$^{+6}_{-1}$\%), type 2
AGNs (AGN2) (50$^{+6}_{-11}$\%) and star-forming galaxy-like AGNs (SF)
(17$^{+9}_{-6}$\%) (Section~\ref{seds}). 

\item {\bf Average SEDs} We derive average SEDs, from hard X-ray to mid-IR 
wavelengths, for each AGN class (Section~\ref{agn_classes_seds}). AGN1s show
AGN-dominated SEDs from optical to infrared wavelengths, and mostly
unabsorbed X-ray spectra. Their optical spectra show a broad range of
extinction, up to \av=0.55, indicating that the AGN1 class includes some
reddened AGNs. In AGN2s, hot dust emission associated with the AGN is
detected at mid-infrared wavelengths, and emission from the host galaxy
dominates in the optical. They show a broad range of X-ray properties, with
the majority being absorbed in the X-rays. Sources classified as SFs do not
show any AGN signature at optical and infrared wavelengths, they are all
absorbed in the X-rays.

\item {\bf Average SEDs $vs$ X-ray luminosity and SED class} The change in
the average SEDs per class as a function of X-ray luminosity is analyzed in
Section~\ref{seds_vs_lx}. The comparison of the average SEDs of each class
shows that the optical-IR SEDs in the AGN2 and SF classes and the IR SEDs in
the AGN1 class redden as the luminosity increases. This reddening is
consistent with a lower relative contribution from the host galaxy to the
near-IR luminosity at larger X-ray luminosities. 

\item {\bf X-ray $vs$ IR luminosity} 
The majority of sources with high X-ray luminosities are AGN1s, while SFs
dominate at lower luminosities. No correlation is observed between the X-ray
or mid-IR luminosity and the X-ray absorption.  

We compare the mid-infrared (3--20~$\mu$m) over absorption-corrected X-ray
luminosity ratio with the X-ray luminosity and with observed values from
various AGN samples in the literature. The luminosity ratios are
characterized by a wide dispersion as the sources in the literature and we
do not find any dependency on the X-ray luminosity. However, we find sources
with significantly lower ratios in the SF class. Even lower ratios are found
for sub-millimeter detected AGNs. These low ratios are explained by either
obscuration in the mid-IR or by lack of hot dust as expected if the
dust is located at a relatively large distance, e.g. 10 pc, from the central
heating source~\citep{ballantyne06}. However, other explanations cannot be
ruled out with the present data.

\item {\bf AGN mid-IR colors} The analysis of AGN red mid-infrared colors as 
measured by IRAC reveals that redness is strongly correlated with the AGN
class, but also with the X-ray and mid-IR luminosities and does not depend
on the X-ray absorption (Section~\ref{agn_colors}). The correlation of the
mid-IR redness with the AGN class is likely due to a decreasing host-galaxy
contribution from SFs to AGN1s. 

\item {\bf Radio luminosity $vs$ X-ray and IR luminosities} Radio
luminosities are compared with the X-ray and mid-infrared luminosities
confirming previously known relationships~\citep{brinkmann00,sopp91} for
AGNs (Section~\ref{radio}). A sub-set of sources shows radio emission in
excess compared with the optical and/or mid-IR flux, consistent with begin
radio-intermediate. We propose the use of the radio luminosity to identify
heavily obscured AGNs by estimating the unabsorbed X-ray luminosity in
radio-quiet AGNs and by comparing the estimated flux with the observed X-ray
flux.

\item {\bf AGN identification through SED fitting}
The SED fitting technique presented here succeeds in identifying signatures
in 83\% of the objects in the hard X-ray sample. The remaining
17$^{+9}_{-6}$\% of AGNs are fitted with star-forming galaxy templates (SF
class). A comparison with the spectroscopic classification available for a
small sub-sample indicates that the AGN1 class is 100\% reliable and 62\%
complete. The AGN2 class is 29\% reliable and 58\% complete, and the SF
class is 33\% reliable and 27\% complete. The SED and the spectroscopic
classification agree in 53\% of the cases.

\item {\bf The nature of elusive AGNs} AGNs in the SF class are elusive AGNs
at optical and IR wavelengths, similarly to optically dull AGNs, XBONGs, and
type 3 QSOs. The properties of these sources can be simply explained by
large obscuration in the mid-IR, however it is not clear whether the large
obscuration is an effect of orientation or whether it implies larger
covering factors and/or dust at larger distance from the nucleus. We do not
find any evidence for inefficient accretion in these sources as they can be
powerful X-ray sources.

\item{\bf The missing AGN population} We investigated whether the AGNs still
missing from X-ray surveys and responsible for the bulk of the CXRB at high
energies (10--30~keV) can be detected at radio wavelengths. The expected
radio flux, derived assuming an X-ray luminosity of 10$^{43}$\ergs\ and the
relationship shown in Figure~\ref{lrad_lx}, and $z$=1 is $\sim$ 2.2
~$\mu$Jy. This flux density is lower than the limits reached by the deepest
current radio surveys. Thus, they will still be missing even in radio
surveys.
\end{itemize}

The SED analysis of the AGN sample presented here indicates the existence of
a large fraction of AGNs that would remain unidentified in optical and/or
IR-selected samples (SFs). These elusive AGNs show a broad range of
luminosities and they are mostly X-ray absorbed. The evolution of this
population needs to be constrained and taken into account in all
evolutionary models. In backward evolutionary models, where SEDs are
associated with different types of
sources~\citep{xu01,xu03,silva04,treister05}, the inclusion of AGNs with
SF-like SED will greatly affect the estimated contribution of AGNs at all
wavelengths, especially in the mid-IR where AGN-dominated SEDs are much
redder than SF SEDs. The identification of these objects would be important
in estimating the contribution of accretion energy to the mid-IR background
and source counts~\citep{silva04,treister06,ballantyne06}.

\acknowledgements
M.P. thanks L. Silva for providing help in producing templates with GRASIL,
B. Schulz for providing the PG quasars ISO-PHT-S spectra, H. McCracken for
useful suggestions, and S. Berta, S. H\"onig and F. Owen for stimulating
discussions. This work is based on observations made with the {\it Spitzer
Space Telescope}, which is operated by the Jet Propulsion Laboratory,
California Institute of Technology under NASA contract 1407. Support for
this work, part of the {\it Spitzer Space Telescope} Legacy Science Program,
was provided by NASA through an award issued by the Jet Propulsion
Laboratory, California Institute of Technology under NASA contract 1407. The
INAF members of the team acknowledge financial contribution from contract
ASI-INAF I/023/05/0. M.T. acknowledges financial support from MIUR Cofin
2004-023189-005. OG and JS acknowledge support from the ESA PRODEX
Programme ``XMM-LSS'', and from the Belgian Federal Science Policy Office
for their support. This research makes use of the NASA/IPAC Extragalactic
Database (NED) which is operated by the Jet Propulsion Laboratory,
California Institute of Technology, under contract with the National
Aeronautics and Space Administration. Based on observations obtained with
MegaPrime/MegaCam, a joint project of CFHT and CEA/DAPNIA, at the
Canada-France-Hawaii Telescope (CFHT) which is operated by the National
Research Council (NRC) of Canada, the Institut National des Science de
l'Univers of the Centre National de la Recherche Scientifique (CNRS) of
France, and the University of Hawaii. This work is based in part on data
products produced at TERAPIX and the Canadian Astronomy Data Centre as part
of the Canada-France-Hawaii Telescope Legacy Survey, a collaborative project
of NRC and CNRS.

{\it Facilities:} \facility{XMM-Newton}, \facility{Spitzer}.

\clearpage
\topmargin=2cm
\footskip=0in
\begin{deluxetable}{rcccc r ccc ccc r r c}
\tabletypesize{\scriptsize}
\rotate
\tablecaption{Luminosities, photometric $z$ and classification of the hard X-ray selected sample\label{lum_tab} }
\tablewidth{0pt}
\tablehead{
\colhead{XMDS seq} & \colhead{Source Name} & \colhead{$z_{phot}$} & \colhead{$z_{spec}$} & \colhead{Sp. type} & \colhead{N$^{eff}_{\mathrm H}$} & \colhead{L$_{BX}$} & \colhead{L$^{corr}_{BX}$} & \colhead{L$^{corr}_{HX}$} & \colhead{L$_{1.4GHz}$}      & \colhead{L$_{mid-IR}$} & \colhead{\av} & \colhead{SED type} & \colhead{SED Class} & \colhead{Comment} \\
\colhead{ }       & \colhead{}            & \colhead{}           & \colhead{}            & \colhead{}           & \colhead{10$^{22}$ cm$^{-2}$}   & \multicolumn{3}{c}{erg s$^{-1}$}                                                          & \colhead{erg s$^{-1}$ Hz$^{-1}$} & \colhead{L$_{\odot}$}       & \colhead{}    & \colhead{}         & \colhead{}      & \colhead{}        }
\startdata
      4 & XMDS J022521.0$-$043949  &    0.245 &    0.265\tablenotemark{a} &        2 &      0.757 &    43.29 &    43.42 &    43.22 & $<$29.21 &    43.57 &     0.40 &       Sb &       SF &              \nodata  \\
     12 & XMDS J022544.9$-$043735  &    3.597 &    3.589\tablenotemark{b} &        1 &     29.825 &    45.54 &    45.67 &    45.42 & $<$31.86 &    46.31 &     0.55 &    TQSO1 &     AGN1 &              \nodata  \\
     13 & XMDS J022504.5$-$043707  &    1.103 &  \nodata                  &  \nodata &      3.287 &    44.32 &    44.45 &    44.20 & $<$30.66 &    45.08 &     0.05 &   Mrk231 &     AGN2 &              \nodata  \\
     16 & XMDS J022506.4$-$043621  &    0.850 &  \nodata                  &  \nodata &   $<$0.129 &    43.97 &    43.97 &    43.68 & $<$30.39 &    44.46 &     0.20 &    Sey18 &     AGN2 &              \nodata  \\
     18 & XMDS J022510.6$-$043549  &    1.612 &  \nodata                  &  \nodata &      1.917 &    44.89 &    44.94 &    44.71 &    31.07 &    45.42 &     0.10 &   I19254 &     AGN2 &              \nodata  \\
     29 & XMDS J022521.0$-$043228  &    0.868 &  \nodata                  &  \nodata &      2.663 &    43.79 &    43.93 &    43.68 & $<$30.42 &    44.41 &     0.10 &   Arp220 &       SF &              \nodata  \\
     36 & XMDS J022449.8$-$043026  &    1.043 &  \nodata                  &  \nodata &      1.773 &    44.44 &    44.53 &    44.32 & $<$30.61 &    45.12 &     0.50 &    BQSO1 &     AGN1 &              \nodata  \\
     40 & XMDS J022510.6$-$042928  &    1.955 &  \nodata                  &  \nodata &     27.436 &    44.92 &    45.14 &    44.91 & $<$31.25 &    45.73 &     0.40 &   Mrk231 &     AGN2 &              \nodata  \\
     55 & XMDS J022522.8$-$042648  &    1.135 &    1.029\tablenotemark{a} &        2 &      4.359 &    44.21 &    44.37 &    44.19 &    31.00 &    45.14 &     0.25 &     QSO2 &     AGN2 &              \nodata  \\
     58 & XMDS J022436.2$-$042511  &    0.645 &  \nodata                  &  \nodata &      1.760 &    43.74 &    43.88 &    43.63 &    30.30 &    44.21 &     0.10 &   Mrk231 &     AGN2 &              \nodata  \\
     60 & XMDS J022439.6$-$042401  &    0.203 &    0.478\tablenotemark{a} &        2 &   $<$0.072 &    43.81 &    43.81 &    43.61 &    30.08 &    44.59 &     0.00 &     QSO2 &     AGN2 &              \nodata  \\
     67 & XMDS J022537.0$-$042132  &    0.961 &  \nodata                  &  \nodata &   $<$0.150 &    44.38 &    44.37 &    44.05 & $<$30.52 &    45.18 &     0.15 &    BQSO1 &     AGN1 &              \nodata  \\
     71 & XMDS J022511.9$-$041911  &    1.401 &  \nodata                  &  \nodata &      8.182 &    44.78 &    44.95 &    44.72 & $<$30.91 &    45.54 &     0.55 &   Mrk231 &     AGN2 &              \nodata  \\
     91 & XMDS J022710.0$-$041649  &    1.005 &  \nodata                  &  \nodata &   $<$0.159 &    44.25 &    44.25 &    44.07 & $<$30.57 &    44.27 &     0.30 &    Sey18 &     AGN2 &                    B  \\
        &                          &    0.396 &  \nodata                  &  \nodata &   $<$0.062 &    43.26 &    43.27 &    43.09 & $<$29.61 &    43.64 &     0.15 &  spi1\_4 &       SF &                    B  \\
     94 & XMDS J022643.8$-$041626  &    0.349 &  \nodata                  &  \nodata &      1.229 &    43.44 &    43.70 &    43.46 & $<$29.48 &    44.34 &     0.00 &     QSO2 &     AGN2 &              \nodata  \\
    101 & XMDS J022809.4$-$041524  &    0.874 &  \nodata                  &  \nodata &   $<$0.134 &    44.24 &    44.24 &    44.03 &  \nodata &    43.69 &     0.00 &       Sb &       SF &                No IR  \\
    106 & XMDS J022719.5$-$041407  &    0.387 &  \nodata                  &  \nodata &     10.289 &    42.73 &    43.09 &    42.83 &    30.20 &    44.11 &     0.55 &   I22491 &     AGN2 &              \nodata  \\
    111 & XMDS J022735.6$-$041317  &    2.015 &  \nodata                  &  \nodata &      5.069 &    45.08 &    45.17 &    45.00 &    31.71 &    45.68 &     0.00 &   I19254 &     AGN2 &             B, No IR  \\
        &                          &    2.514 &  \nodata                  &  \nodata &      7.548 &    45.32 &    45.40 &    45.23 &    31.93 &    44.40 &     0.00 &       Sc &       SF &             B, No IR  \\
    112 & XMDS J022809.0$-$041232  &    0.900 &    0.878\tablenotemark{a} &        1 &      0.386 &    45.08 &    45.12 &    44.87 &  \nodata &    45.86 &     0.55 &    TQSO1 &     AGN1 &                No IR  \\
    114 & XMDS J022649.7$-$041240  &    0.375 &  \nodata                  &  \nodata &   $<$0.060 &    43.22 &    43.22 &    42.99 &    29.60 &    43.73 &     0.15 &       Sc &       SF &              \nodata  \\
    118 & XMDS J022649.3$-$041154  &    1.195 &    1.157\tablenotemark{b} &        1 &   $<$0.193 &    44.38 &    44.37 &    44.05 & $<$30.71 &    44.55 &     0.20 &    Sey18 &     AGN2 &              \nodata  \\
    120 & XMDS J022735.7$-$041122  &    1.167 &  \nodata                  &  \nodata &      0.732 &    45.05 &    45.09 &    44.84 &    31.12 &    44.17 &     0.00 &     Sey2 &     AGN2 &             B, No IR  \\
        &                          &    1.078 &  \nodata                  &  \nodata &      0.656 &    44.96 &    45.00 &    44.76 &    31.03 &    44.61 &     0.15 &    Sey18 &     AGN2 &             B, No IR  \\
    124 & XMDS J022659.7$-$041108  &    0.371 &  \nodata                  &  \nodata &      6.730 &    43.09 &    43.40 &    43.14 & $<$29.54 &    43.92 &     0.10 &     QSO2 &     AGN2 &              \nodata  \\
    133 & XMDS J022713.1$-$040912  &    0.723 &  \nodata                  &  \nodata &      1.200 &    43.91 &    44.00 &    43.74 & $<$30.23 &    43.86 &     0.10 &     QSO2 &     AGN2 &                    M  \\
    134 & XMDS J022701.3$-$040912  &    0.755 &  \nodata                  &  \nodata &   $<$0.113 &    43.74 &    43.74 &    43.52 & $<$30.27 &    44.48 &     0.00 &    Sey18 &     AGN2 &              \nodata  \\
    138 & XMDS J022656.0$-$040821  &    0.546 &  \nodata                  &  \nodata &      7.348 &    43.17 &    43.45 &    43.20 & $<$29.94 &    44.19 &     0.00 &     Sey2 &     AGN2 &              \nodata  \\
    139 & XMDS J022727.7$-$040806  &    0.729 &  \nodata                  &  \nodata &   $<$0.108 &    44.02 &    44.02 &    43.74 & $<$30.23 &    44.50 &     0.00 &    Sey18 &     AGN2 &            Bu, No IR  \\
    140 & XMDS J022701.3$-$040751  &    0.235 &    0.220\tablenotemark{a} &        2 &      0.101 &    43.01 &    43.05 &    42.86 & $<$29.03 &    43.53 &     0.25 &       Sd &       SF &              \nodata  \\
    142 & XMDS J022644.1$-$040720  &    0.573 &  \nodata                  &  \nodata &   $<$0.085 &    43.81 &    43.81 &    43.53 &    30.01 &    44.62 &     0.50 &    BQSO1 &     AGN1 &              \nodata  \\
    143 & XMDS J022655.4$-$040650  &    0.374 &  \nodata                  &  \nodata &      2.817 &    42.98 &    43.20 &    42.95 &    29.59 &    43.89 &     0.35 &     QSO2 &     AGN2 &              \nodata  \\
    144 & XMDS J022652.0$-$040556  &    0.864 &  \nodata                  &  \nodata &   $<$0.132 &    44.28 &    44.28 &    44.09 & $<$30.41 &    44.59 &     0.00 &    Sey18 &     AGN2 &              \nodata  \\
    149 & XMDS J022707.2$-$040438  &    0.493 &  \nodata                  &  \nodata &      0.169 &    43.47 &    43.50 &    43.26 & $<$29.83 &    44.31 &     0.55 &    Sey18 &     AGN2 &              \nodata  \\
    161 & XMDS J022700.7$-$042020  &    0.086 &    0.053\tablenotemark{c} &        1 &   $<$0.030 &    42.83 &    42.84 &    42.44 &    27.95 &    43.51 &     0.00 &     QSO2 &     AGN2 &              \nodata  \\
    178 & XMDS J022544.6$-$041936  &    0.059 &  \nodata                  &  \nodata &      0.223 &    41.82 &    41.90 &    41.66 & $<$27.81 &    42.07 &     0.10 &       S0 &       SF &              \nodata  \\
    179 & XMDS J022607.7$-$041843  &    0.328 &    0.495\tablenotemark{a} &        1 &   $<$0.074 &    44.29 &    44.29 &    44.05 & $<$29.84 &    44.68 &     0.45 &    BQSO1 &     AGN1 &              \nodata  \\
    191 & XMDS J022626.5$-$041214  &    0.779 &  \nodata                  &  \nodata &     19.818 &    44.11 &    44.44 &    44.18 & $<$30.30 &    44.98 &     0.55 &     QSO2 &     AGN2 &              \nodata  \\
    197 & XMDS J022539.0$-$040823  &    0.824 &  \nodata                  &  \nodata &      9.338 &    44.02 &    44.27 &    44.02 & $<$30.36 &    44.76 &     0.55 &    Sey18 &     AGN2 &              \nodata  \\
    199 & XMDS J022614.5$-$040738  &    2.411 &  \nodata                  &  \nodata &     37.558 &    45.15 &    45.36 &    45.12 & $<$31.47 &    45.35 &     0.00 &     QSO2 &     AGN2 &                Bu, M  \\
    227 & XMDS J022511.4$-$041916  &    1.446 &  \nodata                  &  \nodata &     11.553 &    44.74 &    44.94 &    44.69 & $<$30.95 &    44.59 &     0.20 &    N6090 &       SF &                    B  \\
        &                          &    1.683 &  \nodata                  &  \nodata &     14.693 &    44.91 &    45.10 &    44.85 & $<$31.10 &    44.79 &     0.50 &    TQSO1 &     AGN1 &                    B  \\
    229 & XMDS J022406.4$-$041830  &    1.128 &  \nodata                  &  \nodata &     54.869 &    44.27 &    44.69 &    44.43 & $<$30.69 &    44.54 &     0.05 &    N6090 &       SF &              \nodata  \\
    232 & XMDS J022449.2$-$041800  &    0.581 &  \nodata                  &  \nodata &   $<$0.086 &    43.75 &    43.75 &    43.46 & $<$30.00 &    44.01 &     0.00 &    Sey18 &     AGN2 &              \nodata  \\
    233 & XMDS J022456.0$-$041725  &    1.293 &  \nodata                  &  \nodata &      1.114 &    44.45 &    44.50 &    44.25 & $<$30.83 &    44.83 &     0.55 &    TQSO1 &     AGN1 &                   Bu  \\
    242 & XMDS J022437.8$-$041520  &    1.051 &  \nodata                  &  \nodata &      5.200 &    44.07 &    44.24 &    43.99 & $<$30.61 &    44.14 &     0.55 &  spi1\_4 &       SF &              \nodata  \\
    246 & XMDS J022415.6$-$041416  &    2.106 &  \nodata                  &  \nodata &      2.382 &    45.18 &    45.22 &    44.97 & $<$31.33 &    45.84 &     0.50 &     QSO1 &     AGN1 &              \nodata  \\
    253 & XMDS J022451.9$-$041209  &    1.686 &  \nodata                  &  \nodata &   $<$0.341 &    44.79 &    44.78 &    44.55 &    31.35 &    46.08 &     0.15 &    TQSO1 &     AGN1 &              \nodata  \\
    255 & XMDS J022408.4$-$041149  &    1.929 &  \nodata                  &  \nodata &     30.718 &    44.70 &    44.94 &    44.69 &    31.26 &    45.10 &     0.05 &   Arp220 &       SF &                    B  \\
        &                          &    0.357 &  \nodata                  &  \nodata &      4.156 &    42.90 &    43.15 &    42.91 & $<$29.50 &    43.16 &     0.10 &       Sd &       SF &                    B  \\
    258 & XMDS J022447.4$-$041049  &    2.628 &  \nodata                  &  \nodata &     21.429 &    45.08 &    45.23 &    44.98 & $<$31.55 &    45.55 &     0.00 &     QSO1 &     AGN1 &              \nodata  \\
    270 & XMDS J022449.2$-$040841  &    0.958 &  \nodata                  &  \nodata &      2.446 &    43.91 &    44.03 &    43.78 & $<$30.52 &    44.27 &     0.00 &    Sey18 &     AGN2 &              \nodata  \\
    271 & XMDS J022509.5$-$040836  &    2.042 &  \nodata                  &  \nodata &   $<$0.471 &    44.99 &    44.98 &    44.74 & $<$31.30 &    45.84 &     0.40 &    BQSO1 &     AGN1 &              \nodata  \\
    272 & XMDS J022501.6$-$040752  &    0.797 &  \nodata                  &  \nodata &   $<$0.120 &    44.10 &    44.11 &    43.89 & $<$30.33 &    44.98 &     0.25 &    BQSO1 &     AGN1 &              \nodata  \\
    279 & XMDS J022421.3$-$040607  &    0.260 &  \nodata                  &  \nodata &      1.172 &    42.77 &    42.94 &    42.69 & $<$29.19 &    44.00 &     0.35 &     QSO2 &     AGN2 &              \nodata  \\
    280 & XMDS J022417.9$-$040606  &    1.633 &  \nodata                  &  \nodata &     16.445 &    44.59 &    44.80 &    44.55 & $<$31.07 &    44.65 &     0.20 &    TQSO1 &     AGN1 &              \nodata  \\
    281 & XMDS J022503.2$-$040538  &    0.930 &  \nodata                  &  \nodata &      0.061 &    44.61 &    44.61 &    44.36 & $<$30.49 &    44.57 &     0.00 &    Sey18 &     AGN2 &              \nodata  \\
    282 & XMDS J022452.1$-$040518  &    0.189 &  \nodata                  &  \nodata &   $<$0.041 &    42.95 &    42.96 &    42.66 & $<$28.88 &    43.69 &     0.20 &     QSO1 &     AGN1 &              \nodata  \\
    288 & XMDS J022421.2$-$040351  &    0.566 &  \nodata                  &  \nodata &      1.298 &    43.61 &    43.73 &    43.48 & $<$29.97 &    43.36 &     0.35 &spi2a\_12 &       SF &              \nodata  \\
    291 & XMDS J022452.0$-$040258  &    0.269 &  \nodata                  &  \nodata &     54.378 &    42.71 &    43.40 &    43.14 &    29.69 &    43.90 &     0.00 &     QSO2 &     AGN2 &                    M  \\
    330 & XMDS J022333.0$-$041525  &    2.271 &  \nodata                  &  \nodata &     71.370 &    44.73 &    45.03 &    44.78 &  \nodata &    45.60 &     0.00 &     QSO2 &     AGN2 &              \nodata  \\
    351 & XMDS J022356.5$-$041105  &    0.927 &  \nodata                  &  \nodata &      0.679 &    44.06 &    44.10 &    43.86 &  \nodata &    44.25 &     0.00 &    Sey18 &     AGN2 &              \nodata  \\
    359 & XMDS J022325.3$-$040922  &    1.154 &  \nodata                  &  \nodata &     17.337 &    43.93 &    44.20 &    43.95 &  \nodata &    44.26 &     0.00 &  spi1\_4 &       SF &              \nodata  \\
    403 & XMDS J022742.1$-$043607  &    3.493 &  \nodata                  &  \nodata &     32.273 &    45.46 &    45.60 &    45.35 &    32.38 &    45.39 &     0.20 &    N6090 &       SF &                   Bu  \\
    406 & XMDS J022732.7$-$043544  &    0.713 &  \nodata                  &  \nodata &      7.860 &    44.01 &    44.26 &    44.02 &    30.75 &    44.49 &     0.55 &    TQSO1 &     AGN1 &              \nodata  \\
    414 & XMDS J022726.3$-$043327  &    3.666 &  \nodata                  &  \nodata &     10.703 &    45.50 &    45.56 &    45.31 & $<$31.88 &    46.34 &     0.10 &    BQSO1 &     AGN1 &              \nodata  \\
    416 & XMDS J022812.2$-$043230  &    1.668 &  \nodata                  &  \nodata &   $<$0.335 &    44.94 &    44.94 &    44.73 &  \nodata &    45.00 &     0.05 &    BQSO1 &     AGN1 &                No IR  \\
    420 & XMDS J022729.2$-$043225  &    2.357 &    2.290\tablenotemark{b} &        1 &   $<$0.577 &    45.14 &    45.13 &    44.89 & $<$31.41 &    46.20 &     0.00 &     QSO1 &     AGN1 &              \nodata  \\
    427 & XMDS J022758.6$-$043112  &    0.859 &  \nodata                  &  \nodata &   $<$0.131 &    43.91 &    43.91 &    43.68 & $<$30.40 &    44.80 &     0.00 &   I19254 &     AGN2 &                No IR  \\
    430 & XMDS J022737.1$-$043031  &    0.760 &  \nodata                  &  \nodata &      0.537 &    43.93 &    43.98 &    43.74 & $<$30.28 &    43.90 &     0.40 &    Sey18 &     AGN2 &              \nodata  \\
    438 & XMDS J022756.3$-$042905  &    0.375 &  \nodata                  &  \nodata &     12.691 &    42.76 &    43.15 &    42.89 & $<$29.55 &    43.58 &     0.00 &   Arp220 &       SF &                No IR  \\
    439 & XMDS J022746.0$-$042853  &    1.988 &    1.368\tablenotemark{d} &        3 &     13.195 &    44.15 &    44.36 &    44.12 &    30.93 &    44.03 &     0.20 &       Sd &       SF &              \nodata  \\
    440 & XMDS J022748.8$-$042820  &    2.574 &  \nodata                  &  \nodata &      0.509 &    45.43 &    45.42 &    45.22 & $<$31.53 &    46.21 &     0.00 &     QSO1 &     AGN1 &                No IR  \\
    449 & XMDS J022815.2$-$042617  &    2.504 &  \nodata                  &  \nodata &     21.945 &    45.03 &    45.20 &    44.95 &  \nodata &    44.60 &     0.40 &  spi1\_4 &       SF &             B, No IR  \\
        &                          &    2.438 &  \nodata                  &  \nodata &     20.886 &    45.00 &    45.17 &    44.92 &  \nodata &    44.57 &     0.00 &    Sey18 &     AGN2 &             B, No IR  \\
    453 & XMDS J022802.3$-$042546  &    0.568 &  \nodata                  &  \nodata &      5.450 &    43.47 &    43.71 &    43.48 &  \nodata &    44.35 &     0.30 &     QSO2 &     AGN2 &                No IR  \\
    470 & XMDS J022804.5$-$041818  &    1.886 &  \nodata                  &  \nodata &   $<$0.411 &    45.01 &    45.00 &    44.79 &  \nodata &    45.41 &     0.00 &   I19254 &     AGN2 &             B, No IR  \\
        &                          &    0.418 &  \nodata                  &  \nodata &   $<$0.065 &    43.39 &    43.40 &    43.19 &  \nodata &    42.51 &     0.00 &  spi1\_4 &       SF &          B, No IR, M  \\
        &                          &    1.082 &  \nodata                  &  \nodata &   $<$0.176 &    44.41 &    44.41 &    44.20 &  \nodata &    43.88 &     0.00 &   I22491 &     AGN2 &             B, No IR  \\
    487 & XMDS J022643.6$-$043317  &    0.489 &    0.308\tablenotemark{a} &        3 &      5.507 &    42.95 &    43.25 &    43.00 &    29.76 &    44.29 &     0.20 &  spi1\_4 &       SF &                    M  \\
    498 & XMDS J022629.2$-$043057  &    1.903 &    2.031\tablenotemark{a} &        1 &   $<$0.466 &    45.41 &    45.40 &    45.20 & $<$31.29 &    45.82 &     0.00 &     QSO1 &     AGN1 &              \nodata  \\
    503 & XMDS J022649.3$-$042920  &    0.723 &    0.634\tablenotemark{d} &        3 &      4.310 &    43.62 &    43.83 &    43.58 & $<$30.09 &    44.38 &     0.00 &     QSO2 &     AGN2 &              \nodata  \\
    505 & XMDS J022649.0$-$042745  &    0.084 &    0.327\tablenotemark{a} &        2 &   $<$0.054 &    43.07 &    43.08 &    42.86 & $<$29.42 &    44.01 &     0.00 &    Sey18 &     AGN2 &              \nodata  \\
    521 & XMDS J022658.8$-$042321  &    1.754 &    1.325\tablenotemark{d} &        3 &      2.550 &    44.84 &    44.93 &    44.74 &    30.93 &    44.89 &     0.05 &     Sey2 &     AGN2 &              \nodata  \\
    523 & XMDS J022622.1$-$042221  &    1.586 &    2.006\tablenotemark{b} &        1 &   $<$0.456 &    45.40 &    45.39 &    45.11 & $<$31.28 &    45.96 &     0.00 &    BQSO1 &     AGN1 &              \nodata  \\
    551 & XMDS J022342.0$-$043533  &    1.128 &  \nodata                  &  \nodata &      1.723 &    44.44 &    44.52 &    44.28 &  \nodata &    45.12 &     0.35 &     QSO1 &     AGN1 &              \nodata  \\
    561 & XMDS J022424.1$-$043228  &    1.678 &  \nodata                  &  \nodata &   $<$0.338 &    45.05 &    45.05 &    44.72 & $<$31.10 &    45.79 &     0.00 &    BQSO1 &     AGN1 &              \nodata  \\
    564 & XMDS J022350.7$-$043157  &    0.224 &  \nodata                  &  \nodata &   $<$0.044 &    42.71 &    42.72 &    42.51 &  \nodata &    43.57 &     0.00 &    Sey18 &     AGN2 &              \nodata  \\
    565 & XMDS J022356.8$-$043115  &    1.051 &  \nodata                  &  \nodata &      0.774 &    44.26 &    44.31 &    44.06 &  \nodata &    43.98 &     0.55 &  spi1\_4 &       SF &              \nodata  \\
    567 & XMDS J022432.4$-$043036  &    0.588 &  \nodata                  &  \nodata &   $<$0.087 &    43.85 &    43.86 &    43.59 & $<$30.01 &    44.21 &     0.10 &    BQSO1 &     AGN1 &              \nodata  \\
    571 & XMDS J022330.2$-$043004  &    2.404 &    2.666\tablenotemark{a} &        1 &      5.506 &    45.40 &    45.45 &    45.21 &  \nodata &    46.31 &     0.40 &     QSO1 &     AGN1 &              \nodata  \\
    577 & XMDS J022438.9$-$042705  &    0.188 &    0.252\tablenotemark{c} &        2 &   $<$0.047 &    43.69 &    43.70 &    43.39 &    29.51 &    44.31 &     0.10 &    Sey18 &     AGN2 &              \nodata  \\
    578 & XMDS J022350.7$-$042703  &    1.033 &  \nodata                  &  \nodata &   $<$0.165 &    44.22 &    44.22 &    43.88 &  \nodata &    44.69 &     0.25 &     QSO1 &     AGN1 &              \nodata  \\
    602 & XMDS J022351.2$-$042054  &    0.097 &    0.181\tablenotemark{a} &        2 &      2.836 &    42.41 &    42.67 &    42.42 &  \nodata &    43.64 &     0.00 &      M82 &       SF &              \nodata  \\
    626 & XMDS J022326.0$-$043534  &    1.149 &  \nodata                  &  \nodata &      1.191 &    44.63 &    44.69 &    44.45 &  \nodata &    44.91 &     0.00 &    N6090 &       SF &              \nodata  \\
    708 & XMDS J022605.3$-$045803  &    1.476 &  \nodata                  &  \nodata &   $<$0.276 &    44.75 &    44.75 &    44.50 & $<$30.97 &    44.77 &     0.00 &    Sey18 &     AGN2 &              \nodata  \\
    709 & XMDS J022606.7$-$045722  &    0.661 &  \nodata                  &  \nodata &   $<$0.098 &    43.91 &    43.91 &    43.60 & $<$30.13 &    44.26 &     0.00 &    Sey18 &     AGN2 &              \nodata  \\
    710 & XMDS J022627.4$-$045710  &    2.320 &  \nodata                  &  \nodata &      0.931 &    45.55 &    45.56 &    45.35 &    32.11 &    45.97 &     0.50 &    TQSO1 &     AGN1 &              \nodata  \\
    718 & XMDS J022615.1$-$045355  &    0.896 &  \nodata                  &  \nodata &      0.900 &    44.08 &    44.14 &    43.94 &    30.64 &    44.23 &     0.45 &  spi1\_4 &       SF &              \nodata  \\
    720 & XMDS J022628.9$-$045252  &    2.027 &  \nodata                  &  \nodata &   $<$0.465 &    45.14 &    45.13 &    44.88 & $<$31.29 &    45.40 &     0.50 &    BQSO1 &     AGN1 &              \nodata  \\
    731 & XMDS J022554.1$-$044921  &    0.534 &  \nodata                  &  \nodata &      2.107 &    43.26 &    43.42 &    43.17 & $<$29.91 &    43.93 &     0.55 &     QSO2 &     AGN2 &              \nodata  \\
    738 & XMDS J022556.1$-$044724  &    0.898 &    1.010\tablenotemark{a} &        1 &   $<$0.160 &    44.49 &    44.49 &    44.25 &    31.92 &    44.66 &     0.00 &    Sey18 &     AGN2 &              \nodata  \\
    739 & XMDS J022617.1$-$044724  &    0.184 &    0.140\tablenotemark{a} &        2 &      2.820 &    42.11 &    42.38 &    42.14 &    29.32 &    43.56 &     0.05 &  spi1\_4 &       SF &                    M  \\
    742 & XMDS J022514.3$-$044659  &    1.615 &    1.924\tablenotemark{a} &        1 &   $<$0.425 &    45.29 &    45.28 &    45.04 &    31.34 &    46.23 &     0.10 &    BQSO1 &     AGN1 &              \nodata  \\
    743 & XMDS J022625.2$-$044647  &    1.556 &  \nodata                  &  \nodata &      2.256 &    44.74 &    44.80 &    44.56 & $<$31.02 &    45.25 &     0.15 &   Mrk231 &     AGN2 &              \nodata  \\
    746 & XMDS J022512.6$-$044633  &    0.219 &  \nodata                  &  \nodata &      5.213 &    43.04 &    43.36 &    43.10 & $<$29.02 &    43.02 &     0.55 &       Sb &       SF &              \nodata  \\
    747 & XMDS J022640.4$-$044606  &    1.248 &  \nodata                  &  \nodata &     23.531 &    44.49 &    44.79 &    44.53 &    31.61 &    45.17 &     0.00 &   I19254 &     AGN2 &                    B  \\
        &                          &    1.248 &  \nodata                  &  \nodata &     23.531 &    44.49 &    44.79 &    44.53 &    31.61 &    45.17 &     0.00 &   I19254 &     AGN2 &                    B  \\
    748 & XMDS J022610.9$-$044550  &    2.886 &  \nodata                  &  \nodata &     12.547 &    45.32 &    45.42 &    45.17 & $<$31.65 &    45.79 &     0.40 &    TQSO1 &     AGN1 &              \nodata  \\
    755 & XMDS J022600.1$-$044412  &    0.730 &  \nodata                  &  \nodata &   $<$0.108 &    44.06 &    44.06 &    43.83 & $<$30.24 &    44.59 &     0.10 &   Mrk231 &     AGN2 &              \nodata  \\
    760 & XMDS J022531.4$-$044210  &    0.833 &    1.228\tablenotemark{d} &        3 &    150.000\tablenotemark{f} & 45.31 & 46.21 & 45.96 & $<$30.78 & 44.41 &   0.40 &    Sey18 &     AGN2 &              \nodata  \\
    779 & XMDS J022321.8$-$045740  &    0.637 &  \nodata                  &  \nodata &     16.471 &    43.93 &    44.28 &    44.02 &  \nodata &    45.26 &     0.00 &   Mrk231 &     AGN2 &              \nodata  \\
    780 & XMDS J022332.0$-$045740  &    0.963 &  \nodata                  &  \nodata &   $<$0.151 &    44.31 &    44.31 &    44.07 &  \nodata &    45.04 &     0.00 &    Sey18 &     AGN2 &              \nodata  \\
    782 & XMDS J022326.3$-$045708  &    0.839 &    0.826\tablenotemark{a} &        1 &   $<$0.125 &    44.12 &    44.12 &    43.98 &  \nodata &    44.72 &     0.00 &    Sey18 &     AGN2 &              \nodata  \\
    787 & XMDS J022317.9$-$045527  &    0.981 &  \nodata                  &  \nodata &      9.800 &    43.91 &    44.14 &    43.90 &  \nodata &    43.69 &     0.15 &    Sey18 &     AGN2 &              \nodata  \\
    788 & XMDS J022353.7$-$045510  &    0.958 &  \nodata                  &  \nodata &      0.085 &    44.58 &    44.58 &    44.36 &  \nodata &    44.68 &     0.40 &     QSO1 &     AGN1 &              \nodata  \\
    789 & XMDS J022329.1$-$045452  &    0.646 &    0.604\tablenotemark{a} &        1 &   $<$0.089 &    43.84 &    43.84 &    43.56 &  \nodata &    44.44 &     0.00 &    Sey18 &     AGN2 &              \nodata  \\
    800 & XMDS J022403.8$-$045120  &    0.874 &  \nodata                  &  \nodata &   $<$0.134 &    44.13 &    44.13 &    43.88 & $<$30.42 &    44.66 &     0.55 &    Sey18 &     AGN2 &              \nodata  \\
    801 & XMDS J022344.4$-$045120  &    0.959 &  \nodata                  &  \nodata &      1.167 &    44.09 &    44.16 &    43.92 &  \nodata &    44.24 &     0.00 &    Sey18 &     AGN2 &              \nodata  \\
    807 & XMDS J022333.0$-$044924  &    2.039 &    2.302\tablenotemark{a} &        1 &      2.460 &    45.08 &    45.12 &    44.87 &  \nodata &    45.53 &     0.00 &    BQSO1 &     AGN1 &              \nodata  \\
    817 & XMDS J022354.5$-$044815  &    2.428 &    2.458\tablenotemark{a} &        1 &   $<$0.657 &    45.23 &    45.21 &    44.91 &  \nodata &    46.67 &     0.00 &     QSO1 &     AGN1 &              \nodata  \\
    820 & XMDS J022319.4$-$044732  &    0.640 &    0.293\tablenotemark{a} &        2 &      0.185 &    43.18 &    43.23 &    43.03 &  \nodata &    44.27 &     0.00 &     QSO2 &     AGN2 &              \nodata  \\
    825 & XMDS J022330.6$-$044633  &    1.735 &  \nodata                  &  \nodata &      3.157 &    45.05 &    45.12 &    44.84 &  \nodata &    45.34 &     0.10 &   Mrk231 &     AGN2 &              \nodata  \\
    828 & XMDS J022318.8$-$044616  &    0.673 &  \nodata                  &  \nodata &   $<$0.100 &    43.75 &    43.75 &    43.53 &  \nodata &    44.28 &     0.05 &     QSO1 &     AGN1 &              \nodata  \\
    840 & XMDS J022330.9$-$044235  &    2.297 &  \nodata                  &  \nodata &     13.666 &    45.05 &    45.19 &    44.94 &  \nodata &    45.39 &     0.55 &   Mrk231 &     AGN2 &                   Bu  \\
    842 & XMDS J022402.4$-$044140  &    0.010 &   0.0433\tablenotemark{e} &        3 &      1.761 &    41.26 &    41.49 &    41.24 &    29.28 &    43.71 &     0.30 &       Sd &       SF &              \nodata  \\
    844 & XMDS J022343.2$-$044105  &    1.493 &  \nodata                  &  \nodata &     60.833 &    44.46 &    44.83 &    44.57 &  \nodata &    45.36 &     0.05 &   I19254 &     AGN2 &                    B  \\
        &                          &    1.857 &  \nodata                  &  \nodata &     86.702 &    44.69 &    45.07 &    44.81 &  \nodata &    45.32 &     0.25 &   I22491 &     AGN2 &                    B  \\
    846 & XMDS J022317.2$-$044035  &    0.765 &    0.842\tablenotemark{a} &        1 &   $<$0.128 &    44.27 &    44.27 &    44.11 &  \nodata &    45.61 &     0.45 &    BQSO1 &     AGN1 &              \nodata  \\
   1197 & XMDS J022720.2$-$045738  &    1.116 &  \nodata                  &  \nodata &     15.507 &    44.16 &    44.42 &    44.18 & $<$30.68 &    44.95 &     0.25 &   I22491 &     AGN2 &              \nodata  \\
   1199 & XMDS J022651.6$-$045714  &    0.290 &    0.331\tablenotemark{a} &        2 &      0.249 &    43.54 &    43.60 &    43.37 & $<$29.43 &    43.85 &     0.00 &     QSO2 &     AGN2 &              \nodata  \\
   1201 & XMDS J022723.4$-$045608  &    0.168 &  \nodata                  &  \nodata &      0.295 &    42.28 &    42.36 &    42.12 & $<$28.77 &    42.39 &     0.00 &    Sey18 &     AGN2 &              \nodata  \\
   1219 & XMDS J022701.6$-$045158  &    2.112 &  \nodata                  &  \nodata &     25.148 &    44.66 &    44.86 &    44.62 &    32.45 &    45.13 &     0.55 &      M82 &       SF &                   Bu  \\
   1226 & XMDS J022711.7$-$045038  &    0.946 &  \nodata                  &  \nodata &      0.215 &    44.81 &    44.82 &    44.59 & $<$30.50 &    44.93 &     0.55 &    BQSO1 &     AGN1 &              \nodata  \\
   1227 & XMDS J022736.8$-$045033  &    0.445 &  \nodata                  &  \nodata &      1.147 &    43.32 &    43.45 &    43.26 &    29.91 &    44.19 &     0.00 &     Sey2 &     AGN2 &              \nodata  \\
   1231 & XMDS J022731.9$-$044957  &    0.741 &  \nodata                  &  \nodata &   $<$0.110 &    43.62 &    43.62 &    43.39 &    30.85 &    44.18 &     0.30 &    Sey18 &     AGN2 &              \nodata  \\
   1236 & XMDS J022729.0$-$044857  &    1.513 &  \nodata                  &  \nodata &     14.984 &    44.31 &    44.52 &    44.28 & $<$30.99 &    45.51 &     0.40 &   I19254 &     AGN2 &              \nodata  \\
   1246 & XMDS J022712.8$-$044636  &    1.446 &  \nodata                  &  \nodata &   $<$0.267 &    44.74 &    44.74 &    44.48 &    32.85 &    46.02 &     0.00 &    BQSO1 &     AGN1 &              \nodata  \\
   1247 & XMDS J022633.1$-$044637  &    1.197 &  \nodata                  &  \nodata &      3.040 &    44.56 &    44.68 &    44.42 & $<$30.75 &    44.91 &     0.00 &     QSO2 &     AGN2 &              \nodata  \\
   1248 & XMDS J022725.4$-$044619  &    0.034 &  \nodata                  &  \nodata &   $<$0.029 &    41.30 &    41.32 &    41.09 & $<$27.31 &    42.60 &     0.00 &       Sa &       SF &                    M  \\
   1252 & XMDS J022716.0$-$044539  &    0.590 &  \nodata                  &  \nodata &   $<$0.087 &    44.07 &    44.08 &    43.82 & $<$30.02 &    44.55 &     0.50 &    BQSO1 &     AGN1 &              \nodata  \\
   1264 & XMDS J022751.3$-$044251  &    1.694 &  \nodata                  &  \nodata &   $<$0.343 &    44.95 &    44.94 &    44.79 & $<$31.11 &    45.42 &     0.10 &    BQSO1 &     AGN1 &              \nodata  \\
   1265 & XMDS J022712.6$-$044221  &    0.232 &    0.205\tablenotemark{a} &        2 &      2.526 &    42.94 &    43.18 &    42.93 &    29.48 &    43.95 &     0.55 &  spi1\_4 &       SF &              \nodata  \\
\enddata                              
\tablecomments{XMDS seq: X-ray catalog sequence number (Paper I; Paper II);
$z_{phot}$: photometric redshift; $z_{spec}$: spectroscopic redshift; Sp. 
type: spectroscopic classification (1: type 1 AGN, 2: type 2 AGN, 3:
star-forming galaxy). N$^{eff}_{\mathrm H}$: effective column density (Paper
II). Upper limits are derived from the minimum value of 2.6$\times$10$^{20}$
\cm2\, corresponding to the Galactic value, when no additional absorption
was required; L$_{BX}$: logarithm of the not corrected for absorption
rest-frame broad-band (0.5--10~keV) luminosity; L$^{corr}_{BX}$: logarithm
of the absorption-corrected rest-frame broad-band (0.5--10~keV) luminosity;
L$^{corr}_{HX}$: logarithm of the absorption-corrected rest-frame hard-band
(2--10~keV) luminosity. All X-ray luminosities have been derived assuming an
absorbed power-law spectrum with photon index $\Gamma$=1.7 and column
density N$^{eff}_{\mathrm H}$. Broad-band X-ray luminosities have been
derived from the broad-band fluxes and hard-band X-ray luminosities have
been derived from the hard-band fluxes (see Paper II). L$_{1.4GHz}$:
logarithm of the monochromatic radio luminosity at 1.4~GHz rest-frame;
L$_{mid-IR}$: logarithm of the integrated mid-IR luminosity in the
3--20~$\mu$m rest-frame wavelength range; \av: optical extinction derived
from the best-fit assuming the prescription reported in~\citet{calzetti00};
SED type: best-fit template (see Figure~\ref{templates}); Class: AGN
spectral classification (SF, AGN1, AGN2); Comment: B indicates a source in
the B sample, Bu marks those in the B sample with unreliable poorly sampled
SED, M refers to a source for which the absolute magnitude criterion for the
photometric redshift determination was relaxed, No IR indicates sources
without SWIRE coverage.}
\tablenotetext{a}{Spectroscopic redshift from Garcet et al., in prep.}
\tablenotetext{b}{Spectroscopic redshift from~\citet{gavignaud06}.}
\tablenotetext{c}{Spectroscopic redshift from~\citet{lacy06a}.}
\tablenotetext{d}{Spectroscopic redshift from~\citet{lefevre05}.}
\tablenotetext{e}{Spectroscopic redshift from NED.}
\tablenotetext{f}{Fixed maximum value of the intrinsic column density.} 
\end{deluxetable}

\begin{deluxetable}{lcrrr}
\tabletypesize{\scriptsize}
\tablecaption{Fraction of AGNs $vs$ mid-IR color redness \label{lacy_fraction}}
\tablewidth{0pt}
\tablehead{
\colhead{Group}               & \colhead{N} & \colhead{\% in R1} & \colhead{\% in R2} &  \colhead{\% in R3}}
\startdata
All                               &  107 (87) &  28 (25) &  32 (33) &  36 (36) \\
\hline
$L_X<10^{43.8}$                   &   25 (24) &  37 (45) &  29 (34) &  11 (13) \\
$10^{43.8}<L_X<10^{44.3}$         &   29 (21) &  37 (32) &  35 (34) &  16 (13) \\
$10^{44.3}<L_X<10^{44.9}$         &   26 (20) &  23 (18) &  26 (28) &  26 (26) \\
$L_X>10^{44.9}$                   &   22 (17) &   3  (5) &   9  (3) &  47 (48) \\
\hline
$L_{mid-IR}<10^{10.6}$            &   24 (17) &  33 (32) &  21 (21) &  18 (13) \\
$10^{10.6}<L_{mid-IR}<10^{11.0}$  &   29 (23) &  47 (50) &  35 (31) &   8 (10) \\
$10^{11.0}<L_{mid-IR}<10^{11.2}$  &   25 (23) &  17 (14) &  35 (41) &  21 (26) \\
$L_{mid-IR}>10^{11.2}$            &   24 (19) &   3  (5) &   9  (7) &  53 (52) \\
\hline
Normal                            &    0  (0) &   0  (0) &   0  (0) &   0  (0) \\
SF                                &   14  (7) &  30 (27) &   9  (0) &   5  (3) \\
AGN2                              &   55 (45) &  63 (64) &  62 (69) &  39 (35) \\
AGN1                              &   33 (30) &   7  (9) &  29 (31) &  55 (61) \\
\hline
\nh$<10^{21.20}$                  &   25 (24) &  33 (41) &  32 (38) &  11 (13) \\
$10^{21.20}<$\nh$<10^{22.06}$     &   29 (25) &  33 (32) &  32 (38) &  21 (23) \\
$10^{22.06}<$\nh$<10^{22.83}$     &   27 (16) &  30 (23) &  29 (21) &  21 (16) \\
\nh$>10^{22.93}$                  &   21 (17) &   3  (5) &   6  (3) &  47 (48) \\
\enddata
\tablecomments{The reported statistics refers to 107 sources for which IRAC observations are
available. Of these 87 sources are detected in all 4 IRAC bands, for the
remaining 20 sources upper limits were adopted, 4.3, 8.3, 58.5, and 65.7
~$\mu$Jy at 3.6, 4.5, 5.8, and 8.0~$\mu$m, respectively. The values in
parenthesis refer to the sub-sample of 87 sources with detections in all 4
IRAC bands.
$L_X$ is the absorption-corrected luminosity in the 0.5--10~keV rest-frame
interval in \ergs.
$L_{mid-IR}$ is the mid-IR luminosity integrated between 3 and 20 ~$\mu$m
rest-frame in \lsun.
\nh is the intrinsic column density density measured in the X-rays in \cm2.\\
R1, R2, R3 represent three regions defined in color space by the flux
ratios, F(5.8)/F(3.6) and F(8.0)/F(4.0). They are shown in
Figure~\ref{lacy_plots}.
Only sources in the A sample are taken into account.}
\end{deluxetable}

\end{document}